\documentclass[aps,prl,reprint,superscriptaddress,longbibliography]{revtex4-2}

\usepackage{scalerel}
\usepackage{amssymb}
\usepackage{suffix}
\usepackage{float}
\usepackage{mathtools}
\usepackage[utf8]{inputenc}
\usepackage{booktabs}
\usepackage{cases}
\usepackage[multiple]{footmisc}
\usepackage{dcolumn}
\usepackage{color,soul}
\usepackage{rotating}
\usepackage{perpage}
\usepackage{xcolor}
\usepackage{soul}
\usepackage{tikz}
\usepackage[T1]{fontenc}
\usepackage{etoolbox}
\usepackage{graphics}
\usepackage{siunitx}
\usepackage{hyperref}
\usepackage{float}	
\usepackage{collref}
\usepackage{multirow}
\usepackage{mathtools}
\usepackage{bm}
\usepackage{url}

\usepackage{tikz}
\usepackage{tikz-3dplot}
\usepackage{changes}

\newcommand{\red}[1]{\textcolor{black}{#1}}
\newcommand{\mkdel}[1]{}

\newcommand*\rfrac[2]{{}^{#1}\!/_{#2}}
\def\mcirc{\mathbin{\scalerel*{\circ}{j}}}
\def\msquare{\mathord{\scalerel*{\Box}{gX}}}

\begin{document}
	\title{Nonequilibrium phononic first-order phase transition in a driven fermion chain}
	\author{Mohsen Yarmohammadi}
	\email{mohsen.yarmohammadi@utdallas.edu}
	\address{Department of Physics, The University of Texas at Dallas, Richardson, Texas 75080, USA}
	\author{Marin Bukov}
	\email[]{mgbukov@pks.mpg.de}
	\affiliation{Max Planck Institute for the Physics of Complex Systems, N\"othnitzer Str.~38, 01187 Dresden, Germany}
	\author{Michael H. Kolodrubetz}
	\email{mkolodru@utdallas.edu}
	\address{Department of Physics, The University of Texas at Dallas, Richardson, Texas 75080, USA}
	\date{\today}
	
	\begin{abstract}
		We study the direct laser drive of infrared-active phonons that are quadratically coupled to a spinless fermion chain. Feedback is incorporated by phonon dressing of the electronic dispersion, which enables effective non-linearities in the phonon dynamics. We uncover a first-order phase transition in the phononic steady state in which hysteretic effects allow either large or small phonon occupation depending on the drive protocol. We discuss the implications of these findings for probing phase transitions in real driven materials. 
	\end{abstract}
	\maketitle
	
	{\allowdisplaybreaks
		\textit{Introduction.}---Despite a considerable ongoing effort to understand quantum systems, uncovering nonequilibrium phenomena without an apparent equilibrium analogue is one of the outstanding scientific challenges in modern condensed matter physics. Recent developments in the ultrafast dynamics of quantum materials have attracted interest in exploring novel nonequilibrium many-body phenomena~\cite{Novelli2014,DalConte2015,doi:10.1080/00018732.2016.1194044,RevModPhys.83.471,doi:10.1063/PT.3.1717,Basov2017,RevModPhys.93.041002}. Interacting light-matter systems with competing interactions emerged as a useful experimental platform since they allow to study the interplay between coupling to an environment and external drives~\cite{RevModPhys.85.299,PhysRevA.98.023617,Lambert2019}. Moreover, they play a central role in the study of photovoltaics~\cite{Gratzel2001,Wright2016}, light-induced
		phase transitions~\cite{Rini2007,Mitrano2016}, and laser processing~\cite{Malinauskas2016}. 
		\vspace*{-0.05cm}
		
		Direct laser driving of an infrared (IR)-active phonon is the main light-matter interaction channel; it couples to the electron degree of freedom by modifying its environment. In a typical setup, phonons couple to the electron number operator in a nonlinear manner~\cite{Mankowsky_2016}. Over the last decade, a considerable effort has been devoted to studying the role of electron-phonon coupling on the dynamical properties of high-temperature cuprate superconductors~\cite{PhysRevLett.87.077001,PhysRevB.71.094508} and Mott or charge-density-wave insulators~\cite{PhysRevLett.97.067402,PhysRevLett.105.187401,Hellmann2012,Rohwer2011}. For a driven electron-phonon system to avoid heating (e.g., with the goal to store and transmit information), the latter needs to be counterbalanced by dissipation from undriven phonon modes. This has been studied in spin lattices~\cite{PhysRevA.91.051601,PhysRevA.94.033801,yarmohammadi2020dynamical}, coupled quantum-electrodynamics cavities and circuits~\cite{PhysRevLett.116.143603,PhysRevLett.110.163605,PhysRevA.90.023827}, lattice Rydberg atoms~\cite{PhysRevA.84.031402,PhysRevLett.113.210401,PhysRevA.97.053616}, driven-dissipative superfluids~\cite{PhysRevLett.116.235302}, nonlinear photonic modes~\cite{PhysRevA.95.043826,PhysRevA.96.043809}, etc. Although dissipation is widely captured in the dynamics of local observables, a detailed understanding of the process is still an outstanding challenge. 
		\vspace*{-0.05cm}
		
		Studies of interacting systems, exposed to an external laser field and coupled to a thermal bath, have until recently mostly focused on the ``bare'' dispersion of degrees of freedom in dissipation processes. However, it remains to be understood how the backaction of transient excitations can be incorporated into the dissipation process. To the best of our knowledge, this has not been addressed so far in phononically driven materials. This mainly requires a deep scan of the ``dressed'' electronic dispersion through a time-dependent damping rate, which results in a novel physical insight. 
		\begin{figure}[t!]
			\centering
			\includegraphics[width=0.8\linewidth]{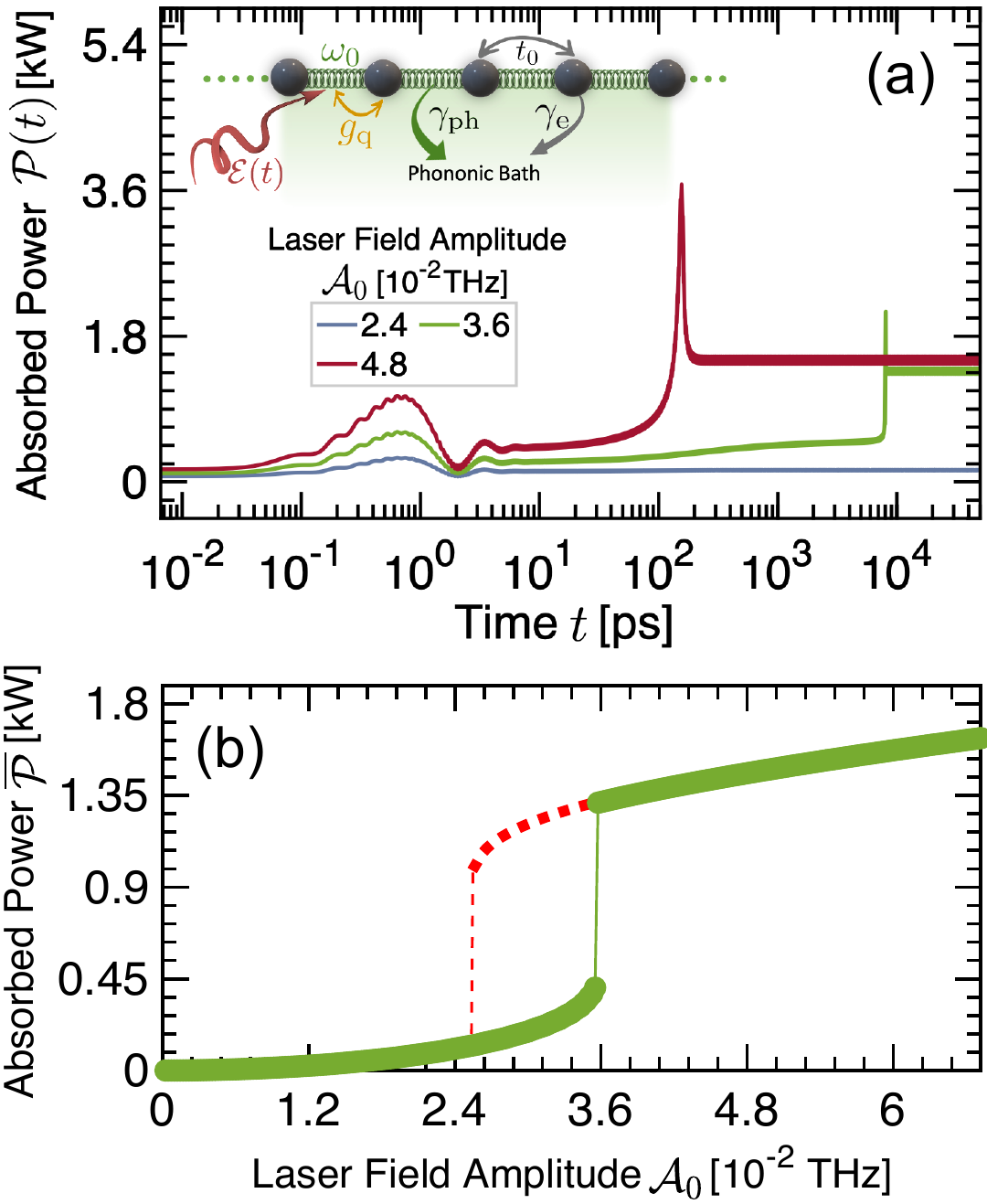}
			\setlength\abovecaptionskip{3pt}
			\caption{\textbf{Nonequilibrium first-order phase transition in a phononically driven fermion chain.} (a) Drive-period-averaged time evolution of absorbed power for a driven-dissipative chain of spinless fermions, sketched in the inset, starting from the ground state at half filling. The system reaches a nonequilibrium steady state~(NESS) which undergoes a first-order phase transition at laser amplitude $\mathcal{A}_0 \approx 3.6\times 10^{-2}$ THz [solid line in (b)] due to phononic modification of the electronic dispersion. The dashed line in (b) shows a second stable NESS which can be created by modifying either the initial state or drive protocol. The parameters are chosen to be $\omega = 4.44$ THz, $\omega_0 = 4.8$ THz, $g_{\rm q} = 9.6$ THz, $\gamma_{\rm ph} = 0.24$ THz, and $\gamma_{\rm e} = 0.0024$ THz; see text for details.}
			\label{f1}
		\end{figure}
		
		In this Letter, we show that a driven-dissipative fermion chain with a dressed dispersion, exhibits a \textit{dynamical first-order phase transition} due to local depopulation of the electrons in the spirit of quadratic electron-phonon coupling~(QEPC). One experimental signature of the transition is a sharp spike in the long time-evolution of the absorbed power, accompanied by a discontinuity in its value in the long-time nonequilibrium steady state~(NESS), as shown in Fig.~\ref{f1}. We analyze and identify accessible parameter regimes to provide a clear path for experimental verification.
		
		\textit{Model.}---Consider an infinite half-filled chain of spinless fermions with periodic boundary conditions, as illustrated in Fig.~\ref{f1}(a), inset. We drive the chain by a continuous field coupled to the IR-active optical phonons, which illuminates the entire system until it reaches a NESS. The model Hamiltonian reads~\cite{PhysRevB.103.054304,Kennes2017,Sous2021}
		\setlength{\abovedisplayskip}{3pt}	\setlength{\belowdisplayskip}{1pt}\begin{equation}\label{eq_1}
			\begin{aligned}
				\mathcal{H}(t)& = -t_0\sum_\ell \big(c^\dagger_\ell c_{\ell+1} + {\rm H.c.}\big) + \omega_0 \sum_\ell a^\dagger_\ell a_\ell \\ &\hspace{-0.85cm}+ g_{\rm q}  \sum_\ell \big(a^\dagger_\ell + a_\ell\big)^2 (c^\dagger_\ell c_\ell -1/2)+ \mathcal{E}(t) \sum_\ell \big(a^\dagger_\ell + a_\ell\big) ,
			\end{aligned}
		\end{equation}where $c^\dagger_\ell~(c_\ell)$ and $a^\dagger_\ell~(a_\ell)$ are respectively the electron and phonon creation~(annihilation) operators at lattice site $\ell$; $t_0$ is a constant hopping amplitude setting the energy scale of our model, $\omega_0$ is the optical phonon frequency, $c^\dagger_\ell c_\ell - 1/2$ is the electron number operator relative to half filling, and $g_{\rm q}$ is the strength of QEPC. In the laser-phonon coupling, the laser field is described by $\mathcal{E}(t) = \mathcal{A}_0 \cos(\omega\,t)$ with frequency $\omega$ and amplitude $\mathcal{A}_0$. In momentum space, we have the bare electronic dispersion $\omega_k=-2t_0\cos k$ and a $k$-independent phonon frequency; the full $k$-space Hamiltonian and details of the model are provided in Sec.~S1 of the Supplemental Material~\cite{SM}. 
		
		We focus our simulations on the dynamics of the approximate dispersionless phonon since, in most materials, the average $\omega_0$ is much larger than the phonon bandwidth. Due to the relatively long wavelength of the drive phonons compared to the lattice spacing, the phonon response is dominated by the zero-momentum mode, $a_0$. We also neglect linear and higher electron-phonon coupling effects, since in target centrosymmetric structures~\cite{Kennes2017,Sous2021} the dominant coupling is quadratic; the contribution of linear electron-phonon coupling is negligible due to a vanishing average of linear oscillations. For comparison to experiments on materials with nonlinear electron-phonon coupling, we choose to consider a representative hopping energy $t_0 = 10\, \text{meV} \approx 2.4\, \text{THz}$, which can also be chosen to be similar to (Pb,Bi)$_2$Sr$_2$CaCu$_2$O$_8$, YBa$_2$Cu$_3$O$_{6+x}$, and K$_3$C$_{60}$~\cite{PhysRevLett.91.167002,Kennes2017,PhysRevB.95.205111,PhysRevB.98.165138} without qualitatively modifying the results.
		
		The system is coupled to a phononic bath~(given by independent phonon modes not subject to the drive) which allows the formation of a NESS. To model the dynamics of the dissipation process, we use the Lindblad master equation for an arbitrary observable $O(t)$:\setlength{\abovedisplayskip}{3pt}	\setlength{\belowdisplayskip}{1pt}\begin{equation}\label{eq_2} 
  \begin{aligned}
		 \langle \dot{O}\rangle (t) &= i\langle[\mathcal{H},O(t)]\rangle  \\ {} &\hspace{-0.85cm}+  \frac{1}{2}\sum_{\ell} \gamma_{\ell} \Big<\big[\mathcal{L}_{\ell}^{\dagger},O(t)\big]\mathcal{L}_{\ell}  + \mathcal{L}_{\ell}^{\dagger}\big[O(t),\mathcal{L}_{\ell}\big]\Big>\, ,
     \end{aligned}
		\end{equation}where the summation $\ell$ describes a bounded set of operators for a separable (system-bath) Hilbert space. The relevant observables are net electron density $n_{\rm e}(t) = {L}^{-1} \sum_k \langle  c^\dagger_k c_k\rangle (t)$, phonon displacement $q_{\rm ph}(t) =  L^{-1/2}\langle a^\dagger_0  + a_0 \rangle (t)$, momentum $p_{\rm ph}(t) = L^{-1/2} \langle i  (a^\dagger_0  - a_0) \rangle (t)$, occupation $n_{\rm ph}(t) = L^{-1} \langle a^\dagger_0 a_0 \rangle (t)$, as well as squeezing-related observables $\mathcal{Q}_{\rm ph}(t) =  L^{-1} \langle (a^\dagger_0a^\dagger_0 + a_0 a_0) \rangle (t)$ and $\mathcal{P}_{\rm ph}(t) =  L^{-1}\langle i  (a^\dagger_0a^\dagger_0 - a_0 a_0) \rangle (t)$. We consider conventional phenomenological relaxation processes given by conventional local jump operators $\mathcal{L}_{\ell} = a^\dagger_0, a_0$, $c^\dagger_k$, and $c_k$ with rates $\gamma_{\rm ph}\,\mathcal{N}_0$, $\gamma_{\rm ph} [1+\mathcal{N}_0]$, $\gamma_{\rm e} \widetilde{\mathcal{N}}_k$, and $\gamma_{\rm e} (1-\widetilde{\mathcal{N}}_k)$, respectively \cite{lindblad1976,breuer2007theory,PhysRevLett.112.117801,yarmohammadi2020dynamical}, where where $\mathcal{N}_0 = 1/[e^{\omega_0/k_{\rm B}T}-1]$ is the average number of phonons in equilibrium and $\widetilde{\mathcal{N}}_k = 1/[e^{\widetilde \omega_k/k_{\rm B}T}+1]$ is the equilibrium electron filling at momentum $k$.
  
        Crucially, in defining the equilibrium electron occupation, we use the \emph{dressed} electron dispersion $\widetilde{\omega}_k(t) = \omega_k + g_{\rm q} q^2_{\rm ph}(t)$ rather than the bare dispersion. Therefore, as the phonon fluctuations $q^2_{\rm ph}$ become large, they shift the chemical potential of the electrons downward, modifying the NESS electron density in the region where the laser drive couples. As we will see, this dressed dispersion gives rise to a nonlinearity that is crucial in understanding the steady state. Note that, by allowing relaxation of the electron density due to a dressed thermal bath, we are assuming that the electrons are connected to some external reservoir, such as metallic leads or the undriven region of the sample. Moreover, we assume that the chain is in touch with a cooling apparatus~\cite{yarmohammadi2020dynamical} to remediate the heating created by the continuous drive. Although our results are robust to the parameters chosen, we consider the experimentally relevant $\omega_0 = 4.8$ THz, $\gamma_{\rm ph} = 0.24$ THz, and $\gamma_{\rm e} = 0.0024$ THz throughout the text, unless otherwise specified. While certain parameters are rationally related, we have confirmed that resonances are irrelevant to the phenomena we describe below.\begin{figure*}[t]
			\centering
			\includegraphics[width=1\linewidth]{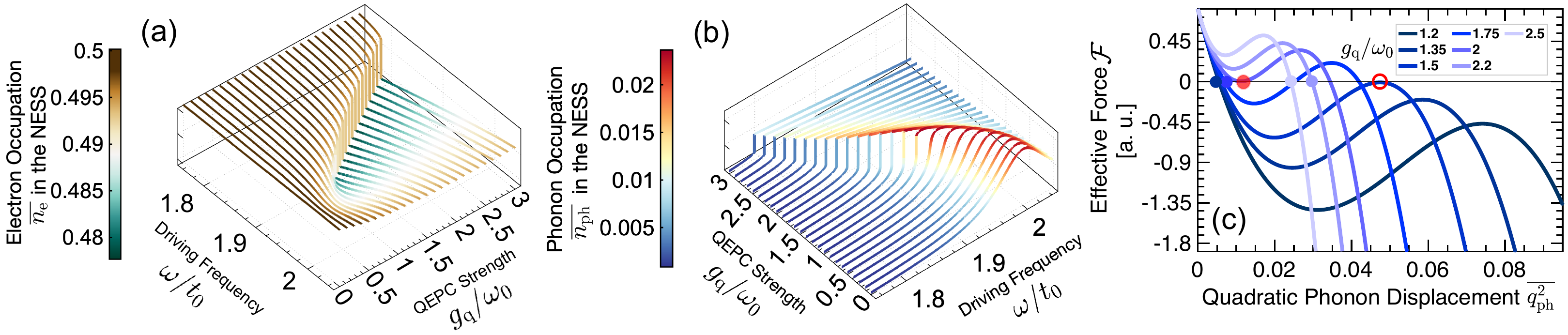}
			\setlength\abovecaptionskip{-12pt}
			\caption{\textbf{Origin of nonequilibrium first-order phase transition.} Dynamical response of dressed (a) electron and (b) phonon occupations in the nonequilibrium steady state for various quadratic electron-phonon couplings and driving frequencies~(axes are swapped in (b) to improve visibility of jumps at the phase transition). Depopulation of the electrons leads to shifting of the phonon resonance, a nonlinearity which enables a first-order phase transition in the NESS at strong $g_{\rm q}$. (c) Nonequilibrium effective force (see text) for various $g_{\rm q}$ at driving frequency $\omega = 4.44$ THz, which is below the bare phonon frequency. Shaded blue dots indicate the numerical steady state. Red dots indicate the two predicted phase transitions (see text), with bistability for $3/2 < g_{\rm q}/\omega_0 < 2$. Parameters are $\omega_0 = 4.8$ THz, $\mathcal{A}_0 = 0.036$ THz, $\gamma_{\rm ph} = 0.24$ THz, and $\gamma_{\rm e} = 0.0024$ THz.} 
			\label{f2}
		\end{figure*}
  
		We initialize the system in its ground state~(electrons at half filling, phonons in the vacuum state), and let it evolve in time following the Lindblad formalism~\cite{lindblad1976,breuer2007theory}. The expectation value of the phonon occupation is proportional to the number of sites $L$, thus, the relative quantum fluctuations which are proportional to $1/\sqrt{L}$ tend to zero in the thermodynamic limit ($L \to \infty$). Additionally, the phonon only couples to electrons via the site-averaged electron density, averaging out on-site electronic fluctuations. We, therefore, employ a mean-field approximation~\cite{yarmohammadi2020dynamical} to decouple the QEPC acting on the phonon and electron, factorizing expectation values according to $\langle A_{\rm ph} B_{\rm e}\rangle \approx \langle A_{\rm ph} \rangle \langle B_{\rm e}\rangle$. The equations then reduce to \begin{subequations} \label{eq_4} 
			\begin{align}  
				\dot{q}_{\rm ph}(t) = {} & + \omega_0 \,p_{\rm ph}(t) -  \frac{\gamma_{\rm ph}}{2} q_{\rm ph}(t)\, ,\label{eq_4a}\\ 
				\dot{p}_{\rm ph}(t) =  {} &- \frac{\widetilde{\omega}^2_0(t)}{\omega_0} q_{\rm ph}(t) \hspace{-0.05cm}-\hspace{-0.05cm}2 \mathcal{E}(t) \hspace{-0.05cm}- \hspace{-0.05cm}\frac{\gamma_{\rm ph}}{2} p_{\rm ph}(t) ,\label{eq_4b}\\
				\dot{n}_{\rm ph}(t) = {} & -\mathcal{E}(t)\, p_{\rm ph}(t) - 2 g_{\rm q} [n_{\rm e}(t)-1/2] \mathcal{P}_{\rm ph}(t) \notag \\ {} &- \gamma_{\rm ph} n_{\rm ph}(t) \, ,\label{eq_s6c}\\
				\dot{\mathcal{Q}}_{\rm ph}(t) = {} & +\Big[\omega_0 + \frac{\widetilde{\omega}^2_0(t)}{\omega_0}\Big] \mathcal{P}_{\rm ph}(t) + 2 \mathcal{E}(t)\, p_{\rm ph}(t)  \notag \\ {} &- \gamma_{\rm ph} \mathcal{Q}_{\rm ph}(t)\,, \label{eq_s6d}\\
				\dot{\mathcal{P}}_{\rm ph}(t) = {}& -\Big[\omega_0 + \frac{\widetilde{\omega}^2_0(t)}{\omega_0}\Big]\mathcal{Q}_{\rm ph}(t)  -2 \mathcal{E}(t)\,q_{\rm ph}(t) \notag \\ &\hspace{-1.5cm} -4 g_{\rm q} \big[2 n_{\rm ph}(t) + \rfrac{1}{L}\big][n_{\rm e}(t)-1/2]-\gamma_{\rm ph} \mathcal{P}_{\rm ph}(t)\, ,\label{eq_s6e}\\
                \dot{n}_{{\rm e},k}(t) = {} &- \gamma_{\rm e} [n_{{\rm e},k}(t)-\widetilde{\mathcal{N}}_k(t)]\, ,\label{eq4f}
			\end{align} 
		\end{subequations}where the electron occupation shifts the effective phonon frequency to $\widetilde{\omega}_0(t) = \omega_0 \sqrt{1{+}4 g_{\rm q} [n_{{\rm e},k}(t) {-} 1/2]/\omega_0}$~\footnote{Note that the phonon relaxation is not modified because it remains gapped throughout the simulations at zero temperature.}.
  
          We consider $T=0$, such that $\widetilde{\mathcal{N}}_k$ equals 1 (filled states) for $|k| < \tilde k_F$ and 0 (empty) otherwise, where $\tilde k_F$ is the Fermi momentum for the dressed dispersion $\widetilde \omega_k$. Despite the complexity of these equations, they can be solved efficiently on a computer; we do so for chain lengths up to $L = 1001$ sites.
  
        As the QEPC $g_{\rm q}$ plays an essential role in determining the dynamics of the model, it is worth noting that the renormalization of the phonon frequency in our model, i.e., $\widetilde{\omega}_0  = \omega_0 \sqrt{1+4 g_{\rm q} [\langle n_{{\rm e}, \ell} \rangle - 1/2]/\omega_0}$ does not lead to instabilities in the simulation, in contrast to other works~\cite{Sous2021,Kennes2017} where the system is unstable for $|g_{\rm q}|>\omega_0/2$. The reason for this is our mean-field-type approximation, which replaces the quantized on-site electron occupation by an average value that only weakly deviates from $1/2$. Crucially, this lack of instability is also physical, suggesting that our results will survive fluctuations. This is because the constraint $|g_{\rm q}|>\omega_0/2$ comes from the assumption of dispersionless Einstein phonons; in reality, there will always be some finite dispersion, causing excitations of the phonons to be spread over at least a few sites. If the minimal spatial dispersion is $\ell$ sites, the instability threshold should be roughly increased by a factor of $\ell$, which brings it well above the value for $g_{\rm q}$ in real materials. Hence, our approximation allows us to simulate the model for arbitrary QEPC strength.
		
		\textit{Results.}---In addition to microscopic quantities such as electron and phonon occupation, we consider the NESS energy flow from the drive into the final stage of dissipation among various degrees of freedom. Of particular importance is the absorbed power 
		\setlength{\abovedisplayskip}{6pt}	\setlength{\belowdisplayskip}{5pt}\begin{equation}\label{eq_2}
			\mathcal{P}(t) = - w a \rho\, \omega_0 \mathcal{E}(t) p_{\rm ph}(t)\, .
		\end{equation}We use parameters from the YBCO sample~\cite{Zhang7}, namely thickness $w = 10$ nm, area $a = 1$ mm$^2$, and molar density $\rho \approx 0.007$ mol.cm$^{-3}$. This power is related to experimentally measurable quantities such as reflectance~\cite{https://doi.org/10.48550/arxiv.2301.08633,https://doi.org/10.48550/arxiv.2301.06425,Dolgirev2022,PhysRevX.12.031008,Henstridge2022,PhysRevLett.127.197002,PhysRevX.11.011055,PhysRevX.10.031028}. 

        Upon quenching on the drive and solving the dynamics, the system evolves into a NESS as seen in Fig.~\ref{f1}(a). For the majority of parameters, it takes around 600 ps to reach the NESS, which is within the current technological capabilities of laser sources in ultrafast experiments. In the presence of a strong QEPC, as laser amplitude $\mathcal{A}_0$ is increased, the smooth transient process of the time-evolved absorbed power suddenly becomes a sharp peak at a critical value of $\mathcal{A}_0 = 0.036$ THz. At this point, observables including the absorbed power undergo a non-analytic change and a \textit{first-order phase transition} occurs, which comes from the depopulation of the electronic modes via feedback from the phonons in the dissipation process, i.e., from $\widetilde{\mathcal{N}}_k(t)$ in Eq.~\eqref{eq4f}. This can also be confirmed by focusing on the NESS in Fig.~\ref{f1}(b), i.e., by averaging an observable $O(t)$ over one drive period, denoted by $\overline{O}$. As we show later, modifying either the initial state or the drive protocol leads to a different NESS, as shown by the dashed red line in Fig.~\ref{f1}(b) and supported by the data in Fig.~\ref{f4}.
		
	 To explain the origin of phase transition and to check its robustness against changing model parameters, we continue by discussing the response of the electrons and phonons to changing drive frequency $\omega$ and QEPC $g_{\rm q}$ [Figs.~\ref{f2}(a) and~\ref{f2}(b)]. In the absence of QEPC, there is a resonant peak at $\omega = \omega_0$, more clearly visible in the phonon occupation. As $g_{\rm q}$ is increased, the peak shifts to a lower frequency due to the depopulation of the electron density via feedback from the phonons. Surprisingly, above a critical value of $g_{\rm q}$, the smooth peak suddenly becomes a sharp jump, suggestive of a first-order phase transition. 
		
    The main origin of the phase transition is dissipation-induced nonlinearity. The dressed electron dispersion $\widetilde{\omega}_k(t) = \omega_k + g_{\rm q} q^2_{\rm ph}(t)$ will yield an average shift of the electron chemical potential given by $g_{\rm q} \overline{q^2_{\rm ph}}$~(note that $\overline{q_{\rm ph}} = 0$). Linearizing the electron density around the Fermi surface via ${L}^{-1} \sum_k \widetilde{\mathcal{N}}_k(t) = \tilde{k}_{\rm F}(t)/\pi$, where $\tilde{k}_{\rm F}(t) = \pi/2 - g_{\rm q} q^2_{\rm ph}(t)/2 t_0$, we predict a steady state electron density of\setlength{\abovedisplayskip}{0pt}	\setlength{\belowdisplayskip}{1pt}\begin{equation}\label{eq_3}
			\overline{n_{\rm e}} = \frac{1}{2} - \frac{g_{\rm q} \overline{q^2_{\rm ph}}}{2 \pi t_0}\, .
		\end{equation}Since the steady state to which the electrons attempt to relax to depends on the phonon state, whose dynamics, in turn, depend on the electron density, this results in effective nonlinearities in the dynamics. This backaction was not considered in previous works~\cite{Kennes2017,Sous2021} because the electron was assumed to relax to its undriven ground state. While reasonable for linear electron-phonon coupling, the presence of a finite $\overline{q^2_{\rm ph}}$ makes this backaction crucial for QEPC~\cite{PhysRevA.98.052129}. 
  
        Correctly accounting for this backaction is our major contribution to the model which, as we have seen, produces significant effects on the dynamics. It is also important to note that $q^2_{\rm ph}(t)$ can be experimentally measured through the intensity of a diffraction peak in femtosecond time-resolved X-rays~\cite{PhysRevLett.102.175503}, enabling another path to measure the effects of this backaction.
  
        To microscopically interpret the observed phase transition, we consider the system near its steady-state; we assume that the phonons synchronize with the drive, with the dominant Fourier component of $q_{\rm ph}(t) = q_1 e^{i\omega t} + \mathrm{c.c.}$ and similarly for $p_{\rm ph}$. Away from the NESS, $q_1$ slowly evolves towards a stationary point, which can be obtained by setting $\dot{q}_1=\dot{p}_1 = 0$. Solving Eqs.~\eqref{eq_4a} and~\eqref{eq_4b} for this Fourier mode, we obtain\begin{subequations}
			\begin{align}
				i \omega q_1 = {} & \omega_0 p_1 - \frac{\gamma_{\rm ph}}{2} q_1\, ,\\
				i \omega p_1 = {} & - \omega_0 q_1 - 4 q_{\rm q} [\overline{n_{\rm e}}-1/2] q_1 - 2 \mathcal{A}_0 - \frac{\gamma_{\rm ph}}{2} q_1\, .
			\end{align}
		\end{subequations}Using Eq.~\eqref{eq_3} and the relation $\overline{q^2_{\rm ph}} = |q_1|^2/2$, this gives a cubic equation \begin{equation}\label{eq_4}
			\begin{aligned}
				\mathcal{F} = {}& -\frac{4 g^4_{\rm q} \omega^2_0}{\pi^2 t^2_0} \,\overline{q^2_{\rm ph}}^3 -  \frac{4 g^2_{\rm q} \omega_0}{\pi t_0} \bigg(\omega^2 - \omega^2_0 + \frac{\gamma^2_{\rm ph}}{4}\bigg) \overline{q^2_{\rm ph}}^2 \\&- \Big(\gamma^2_{\rm ph} \omega^2 + \big(\omega^2 - \omega^2_0\big)^2\Big)\overline{q^2_{\rm ph}} +2 \mathcal{A}^2_0 \omega^2_0,
			\end{aligned}
		\end{equation}with $\mathcal{F}=0$ in the NESS. Perturbing away from the $\mathcal{F}=0$ solution, this cubic equation may be thought of as a generalized force; stable attractors have $d\mathcal{F}/d\overline{q^2_{\rm ph}} < 0$.
  
        As seen in Fig.~\ref{f2}(c), the force curve shifts depending on various parameters and appears to have two separate phase transitions (marked by red circles) from having a unique NESS to a regime with two distinct stable NESSs. The first phase transition~(transparent circle) is visible in our numerical data since, starting from a vacuum state, a jump from small to large $\overline{q^2_{\rm ph}}$ matches our expectations. The second phase transition~(open circle) is not seen in the data, but could be realized via protocols similar to those in Fig.~\ref{f4}. This bifurcation transition with a cubic equation for the effective force is reminiscent of the magnetization in a first-order Ising phase transition. However, we emphasize that our phase transition occurs not in equilibrium, but rather out of equilibrium in the driven NESS.\begin{figure}[t]
		\centering
		\includegraphics[width=0.8\linewidth]{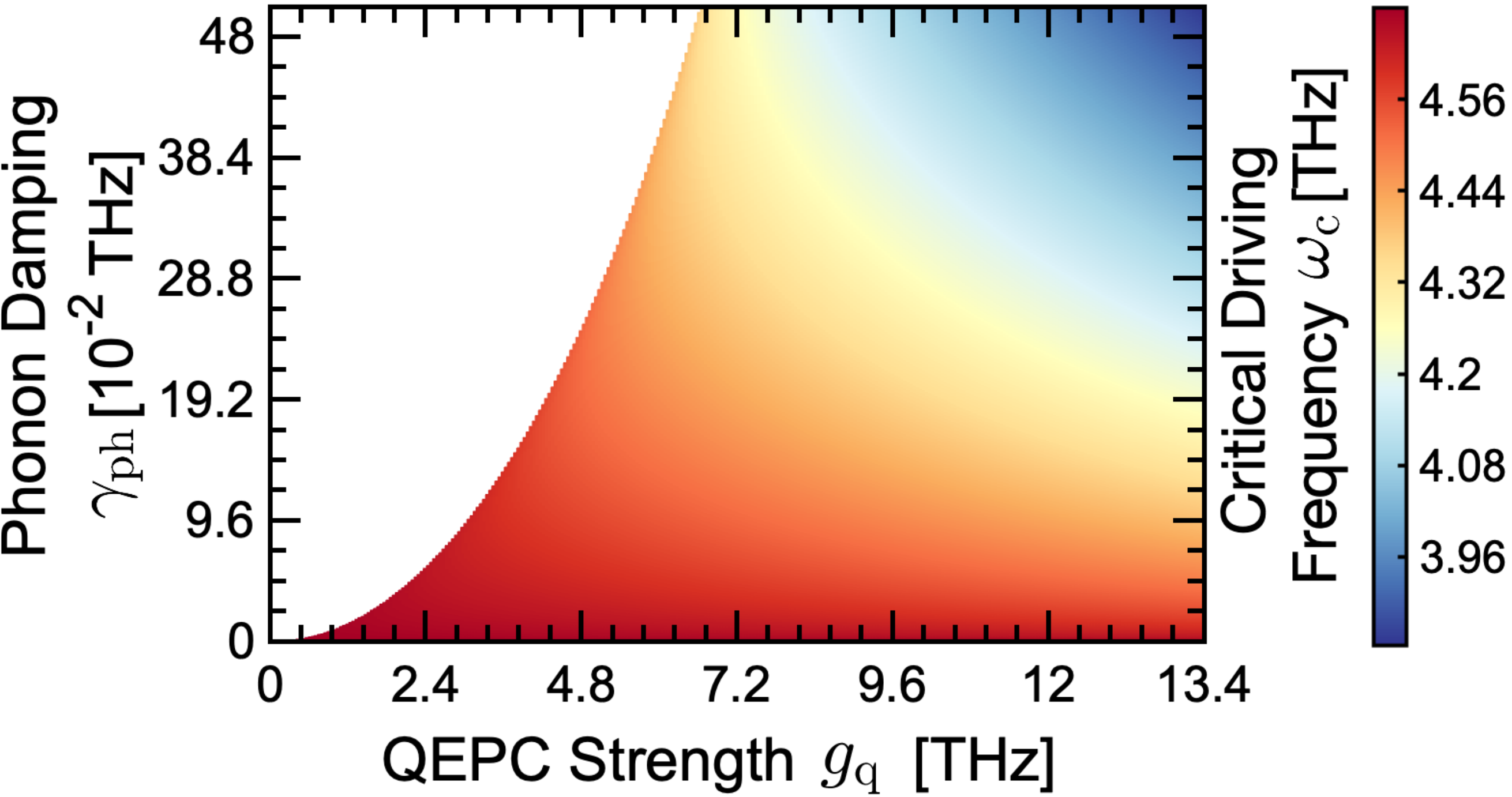}
		\setlength\abovecaptionskip{3pt}
		\caption{\textbf{Dynamical damping-coupling constraint for the phase transition.} Critical driving frequency as a function of phonon damping rate and strength of quadratic electron-phonon coupling at $\omega_0 = 4.8$ THz, $\mathcal{A}_0 = 0.036$ THz, and $\gamma_{\rm e} = 0.0024$ THz. The transition occurs down to arbitrarily small values of $g_{\mathrm q}$ and $\gamma_\mathrm{ph}$. The separatrix, below which dynamical first-order phase transitions occur, is parabolic in the $\gamma_{\rm ph}-g_{\rm q}$ plane.} 
		\label{f3}
	\end{figure}
 
	One may naively conclude from Fig.~\ref{f2} that the phase transition requires extremely large QEPCs, which can be challenging to find experimentally. However, by solving the cubic equation, we find the following analytical expression for the critical driving frequency:\begin{subequations}\label{eq_5}
	    \begin{align}
            \omega_{\rm c}  =  {} &\sqrt{\omega^2_0 - \frac{2\sqrt{3}\gamma_{\rm ph} \omega_0}{3}  + \frac{W}{2}}\, ,\\
            W = {} & \sqrt{\gamma_{\rm ph}^4 - \frac{\pi t_0\tilde{b}}{3 g^2_{\rm q}\omega_0} - 4 \gamma_{\rm ph}^2 \omega_0^2- \frac{96 g^4_{\rm q} \omega^4_0 \mathcal{A}^2_0}{\pi^2 t_0^2 \tilde{b}}}\, ,
     \end{align}
	\end{subequations}where $\tilde{b} = g^2_{\rm q} \omega_0 \gamma_{\rm ph} (3 \gamma_{\rm ph} - 4 \sqrt{3} \omega_0)/\pi t_0$. Importantly, this shows that a phase transition persists down to arbitrarily small $q_{\rm q}$, as seen in the parametric plot of $\gamma_{\rm ph}$ vs. $g_{\rm q}$ in Fig.~\ref{f3}. Intriguingly, even with 1\% damping of phonon energy~($\gamma_{\rm ph} \approx 0.048$ THz) to the phononic bath, governed by the Lindemann criterion~\cite{Lindemann1910}, the system still features the phase transition for moderate QEPC $g_{\rm q} = \omega_0/2 = t_0$. The region in which the transition can be observed increases with the phonon frequency.
	\begin{figure}[t]
		\centering
		\includegraphics[width=0.8\linewidth]{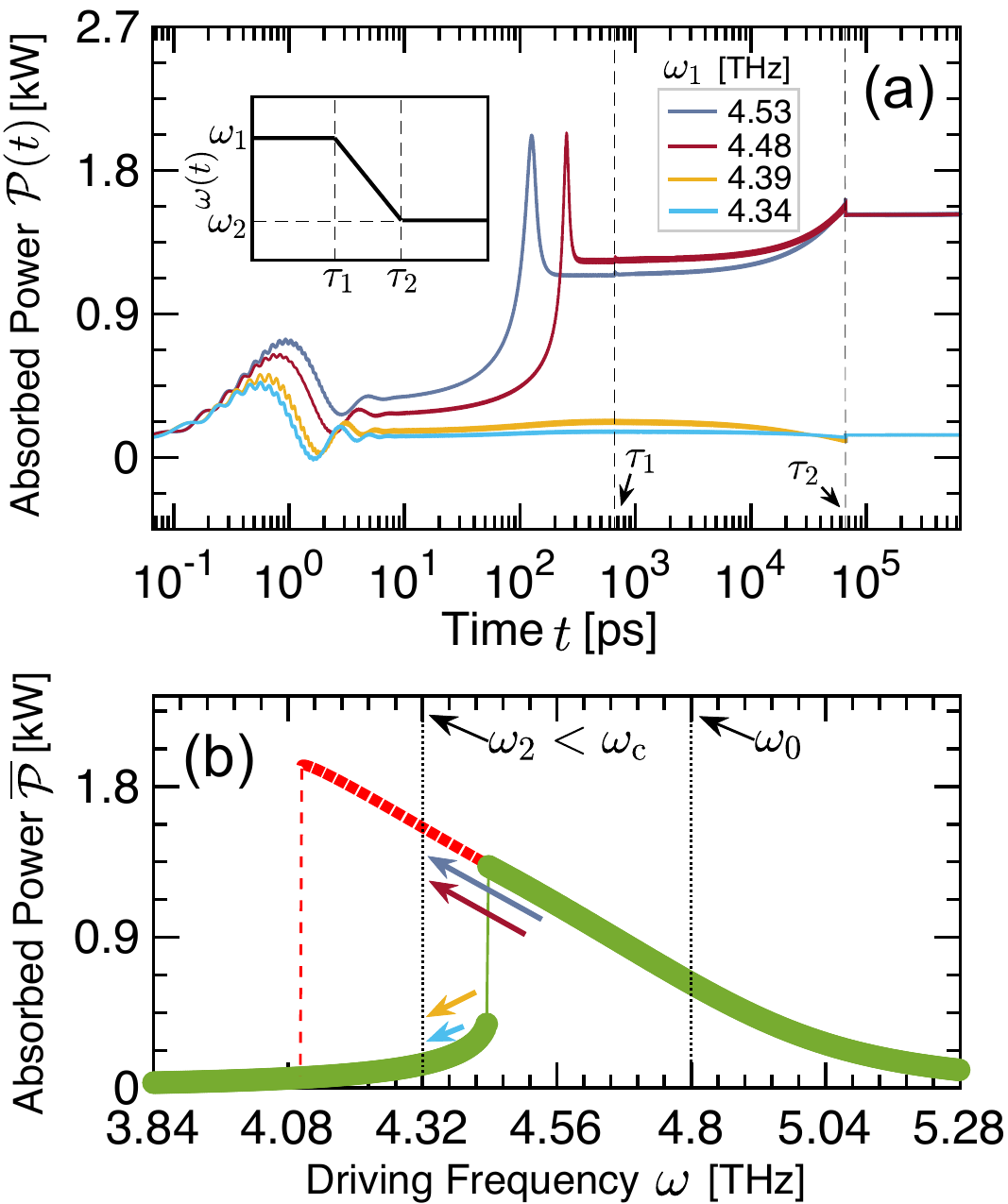}
		\setlength\abovecaptionskip{3pt}
		\caption{\textbf{Chirp protocol for realizing the alternative steady state.} (a) Absorbed power in presence of a linearly chirped electric field $\mathcal{E}(t) = \mathcal{A}_0 \cos\big(\omega(t)\,t\big)$ with $\omega(t) = \omega_1 + [(\omega_2-\omega_1)(t-\tau_1)/(\tau_2-\tau_1)]$ (inset) for fixed $\omega_2 = 4.32$ \rm{THz} < $\omega_c = 4.44$ \rm{THz}. (b) Depending on whether $\omega_1$ is less than or greater than $\omega_c$, the system ends up in a different branch of the steady state. Black arrows point to the positions of the frequencies $\omega_0$ and $\omega_2$ (black dotted vertical lines); color-coded arrows indicate the value of $\omega_1$, cf.~legend in (a). Parameters are $\omega_0 = 4.8$ THz, $\mathcal{A}_0 = 0.036$ THz, $\gamma_{\rm ph} = 0.24$ THz, and $\gamma_{\rm e} = 0.0024$ THz.} 
		\label{f4}
	\end{figure}
	
	The generalized force in Fig.~\ref{f2}(c) suggests that two stable equilibria exist at large $g_{\rm q}$ or small $\omega$, yet only one NESS is seen in our simulation for each value of the parameters. This is the attractor of the equations of motion starting from our initial ground state. To realize the other NESS, a different initial state must be prepared. Taking the cue from hysteresis near equilibrium first-order phase transitions, we suggest that this may be done via slowly ramping one of the parameters, such as the frequency (a ``chirp'' protocol). We consider a linearly chirped electric field $\mathcal{E}(t) {=} \mathcal{A}_0 \cos\big(\omega(t)\,t\big)$ with $\omega(t) {=} \omega_1 {+} (\omega_2{-}\omega_1)(t{-}\tau_1)/(\tau_2{-}\tau_1)$, as shown in Fig.~\ref{f4}(a), inset. Such frequency ramps are important throughout ultrafast spectroscopy~\cite{PhysRevLett.122.204802,PhysRevLett.120.044802}, and can thus be implemented experimentally. Starting with frequency $\omega_1$ at time $\tau_1$ sufficiently large to reach the NESS, we slowly ramp to the final value $\omega_2 < \omega_c$ at time $\tau_2$. If $\omega_1 < \omega_c$ as well, no phase transition is crossed by this ramp and the system simply reaches the original NESS. However, if $\omega_1 > \omega_c$, the system instead stays in the upper NESS with large $\overline{q^2_{\rm ph}}$, as shown in Fig.~\ref{f4}(b). Similar hysteretic preparation of the alternative NESS can be accomplished via slowly ramping other experimental parameters, such as the drive amplitude $\mathcal{A}_0$ in Fig.~\ref{f1}(b).
	
	\textit{Conclusion/Outlook.}---We have uncovered a novel nonequilibrium phase transition of a driven-dissipative fermion chain coupled to a phonon mode. By dynamically tuning the dissipation processes through the backaction of phononic excitations on the electronic dispersion, we find a robust phase transition from having a unique nonequilibrium steady state to having two stable steady states. Since the effect is favored by a quasi-equilibrium electron density~(less than 2\% deviation from a half-filled chain in equilibrium), it should be prevalent in any fermion chain. Our findings motivate future nonequilibrium spectroscopy experiments to seek photo-induced phononic phase transitions in driven quantum materials such as driven superconductors, where the modified phononic steady state may enable a nonequilibrium pathway to controlling superconductivity~\cite{eckhardt2023theory,chattopadhyay2023mechanisms}. Since such strong phononic steady state tends to be inhibited by the feedback mechanism that we consider, nonequilibrium protocols such as the chirp may be required to push these systems into the desired NESS. This phase transition will likely also be possible in the presence of an electromagnetic cavity, which has been used in recent work to control heating in driven materials~\cite{Schlawin2022,doi:10.1126/sciadv.aau6969,LeDe_2022}. 
 
    Finally, we note that these nonequilibrium phase transitions not restricted to 1D, and are in fact more likely in higher dimensions, where mean-field theory and expansions around the Fermi sea are even more well-justified. Indeed, the phenomenon we reported primarily comes from a mean-field treatment of the average density, whose weakly correlated behavior is insensitive to dimensionality up to an overall prefactor in the density of states.
	
	\textit{Acknowledgments.}---This work was performed with support from the National Science Foundation (NSF) through award numbers MPS-2228725 and DMR-1945529 and the Welch Foundation through award number AT-2036-20200401 (MK and MY). Part of this work was performed at the Aspen Center for Physics, which is supported by NSF grant No. PHY-1607611, and at the Kavli Institute for Theoretical Physics, which is supported by NSF grant No. NSF PHY-1748958. This project was funded by The University of Texas at Dallas Office of Research and Innovation through the SPIRe program. M.B. was supported by Marie Skłodowska Curie Grant No.~890711 (until 01.09.2022).
}
	\bibliography{bibliography}

\begin{thebibliography}{64}%
\makeatletter
\providecommand \@ifxundefined [1]{%
 \@ifx{#1\undefined}
}%
\providecommand \@ifnum [1]{%
 \ifnum #1\expandafter \@firstoftwo
 \else \expandafter \@secondoftwo
 \fi
}%
\providecommand \@ifx [1]{%
 \ifx #1\expandafter \@firstoftwo
 \else \expandafter \@secondoftwo
 \fi
}%
\providecommand \natexlab [1]{#1}%
\providecommand \enquote  [1]{``#1''}%
\providecommand \bibnamefont  [1]{#1}%
\providecommand \bibfnamefont [1]{#1}%
\providecommand \citenamefont [1]{#1}%
\providecommand \href@noop [0]{\@secondoftwo}%
\providecommand \href [0]{\begingroup \@sanitize@url \@href}%
\providecommand \@href[1]{\@@startlink{#1}\@@href}%
\providecommand \@@href[1]{\endgroup#1\@@endlink}%
\providecommand \@sanitize@url [0]{\catcode `\\12\catcode `\$12\catcode
  `\&12\catcode `\#12\catcode `\^12\catcode `\_12\catcode `\%12\relax}%
\providecommand \@@startlink[1]{}%
\providecommand \@@endlink[0]{}%
\providecommand \url  [0]{\begingroup\@sanitize@url \@url }%
\providecommand \@url [1]{\endgroup\@href {#1}{\urlprefix }}%
\providecommand \urlprefix  [0]{URL }%
\providecommand \Eprint [0]{\href }%
\providecommand \doibase [0]{https://doi.org/}%
\providecommand \selectlanguage [0]{\@gobble}%
\providecommand \bibinfo  [0]{\@secondoftwo}%
\providecommand \bibfield  [0]{\@secondoftwo}%
\providecommand \translation [1]{[#1]}%
\providecommand \BibitemOpen [0]{}%
\providecommand \bibitemStop [0]{}%
\providecommand \bibitemNoStop [0]{.\EOS\space}%
\providecommand \EOS [0]{\spacefactor3000\relax}%
\providecommand \BibitemShut  [1]{\csname bibitem#1\endcsname}%
\let\auto@bib@innerbib\@empty
\bibitem [{\citenamefont {Novelli}\ \emph {et~al.}(2014)\citenamefont
  {Novelli}, \citenamefont {De~Filippis}, \citenamefont {Cataudella},
  \citenamefont {Esposito}, \citenamefont {Vergara}, \citenamefont {Cilento},
  \citenamefont {Sindici}, \citenamefont {Amaricci}, \citenamefont {Giannetti},
  \citenamefont {Prabhakaran}, \citenamefont {Wall}, \citenamefont {Perucchi},
  \citenamefont {Dal~Conte}, \citenamefont {Cerullo}, \citenamefont {Capone},
  \citenamefont {Mishchenko}, \citenamefont {Gr{\"u}ninger}, \citenamefont
  {Nagaosa}, \citenamefont {Parmigiani},\ and\ \citenamefont
  {Fausti}}]{Novelli2014}%
  \BibitemOpen
  \bibfield  {author} {\bibinfo {author} {\bibfnamefont {F.}~\bibnamefont
  {Novelli}}, \bibinfo {author} {\bibfnamefont {G.}~\bibnamefont
  {De~Filippis}}, \bibinfo {author} {\bibfnamefont {V.}~\bibnamefont
  {Cataudella}}, \bibinfo {author} {\bibfnamefont {M.}~\bibnamefont
  {Esposito}}, \bibinfo {author} {\bibfnamefont {I.}~\bibnamefont {Vergara}},
  \bibinfo {author} {\bibfnamefont {F.}~\bibnamefont {Cilento}}, \bibinfo
  {author} {\bibfnamefont {E.}~\bibnamefont {Sindici}}, \bibinfo {author}
  {\bibfnamefont {A.}~\bibnamefont {Amaricci}}, \bibinfo {author}
  {\bibfnamefont {C.}~\bibnamefont {Giannetti}}, \bibinfo {author}
  {\bibfnamefont {D.}~\bibnamefont {Prabhakaran}}, \bibinfo {author}
  {\bibfnamefont {S.}~\bibnamefont {Wall}}, \bibinfo {author} {\bibfnamefont
  {A.}~\bibnamefont {Perucchi}}, \bibinfo {author} {\bibfnamefont
  {S.}~\bibnamefont {Dal~Conte}}, \bibinfo {author} {\bibfnamefont
  {G.}~\bibnamefont {Cerullo}}, \bibinfo {author} {\bibfnamefont
  {M.}~\bibnamefont {Capone}}, \bibinfo {author} {\bibfnamefont
  {A.}~\bibnamefont {Mishchenko}}, \bibinfo {author} {\bibfnamefont
  {M.}~\bibnamefont {Gr{\"u}ninger}}, \bibinfo {author} {\bibfnamefont
  {N.}~\bibnamefont {Nagaosa}}, \bibinfo {author} {\bibfnamefont
  {F.}~\bibnamefont {Parmigiani}},\ and\ \bibinfo {author} {\bibfnamefont
  {D.}~\bibnamefont {Fausti}},\ }\bibfield  {title} {\bibinfo {title}
  {Witnessing the formation and relaxation of dressed quasi-particles in a
  strongly correlated electron system},\ }\href
  {https://doi.org/10.1038/ncomms6112} {\bibfield  {journal} {\bibinfo
  {journal} {Nature Communications}\ }\textbf {\bibinfo {volume} {5}},\
  \bibinfo {pages} {5112} (\bibinfo {year} {2014})}\BibitemShut {NoStop}%
\bibitem [{\citenamefont {Dal~Conte}\ \emph {et~al.}(2015)\citenamefont
  {Dal~Conte}, \citenamefont {Vidmar}, \citenamefont {Gole{\v{z}}},
  \citenamefont {Mierzejewski}, \citenamefont {Soavi}, \citenamefont {Peli},
  \citenamefont {Banfi}, \citenamefont {Ferrini}, \citenamefont {Comin},
  \citenamefont {Ludbrook}, \citenamefont {Chauviere}, \citenamefont
  {Zhigadlo}, \citenamefont {Eisaki}, \citenamefont {Greven}, \citenamefont
  {Lupi}, \citenamefont {Damascelli}, \citenamefont {Brida}, \citenamefont
  {Capone}, \citenamefont {Bon{\v{c}}a}, \citenamefont {Cerullo},\ and\
  \citenamefont {Giannetti}}]{DalConte2015}%
  \BibitemOpen
  \bibfield  {author} {\bibinfo {author} {\bibfnamefont {S.}~\bibnamefont
  {Dal~Conte}}, \bibinfo {author} {\bibfnamefont {L.}~\bibnamefont {Vidmar}},
  \bibinfo {author} {\bibfnamefont {D.}~\bibnamefont {Gole{\v{z}}}}, \bibinfo
  {author} {\bibfnamefont {M.}~\bibnamefont {Mierzejewski}}, \bibinfo {author}
  {\bibfnamefont {G.}~\bibnamefont {Soavi}}, \bibinfo {author} {\bibfnamefont
  {S.}~\bibnamefont {Peli}}, \bibinfo {author} {\bibfnamefont {F.}~\bibnamefont
  {Banfi}}, \bibinfo {author} {\bibfnamefont {G.}~\bibnamefont {Ferrini}},
  \bibinfo {author} {\bibfnamefont {R.}~\bibnamefont {Comin}}, \bibinfo
  {author} {\bibfnamefont {B.~M.}\ \bibnamefont {Ludbrook}}, \bibinfo {author}
  {\bibfnamefont {L.}~\bibnamefont {Chauviere}}, \bibinfo {author}
  {\bibfnamefont {N.~D.}\ \bibnamefont {Zhigadlo}}, \bibinfo {author}
  {\bibfnamefont {H.}~\bibnamefont {Eisaki}}, \bibinfo {author} {\bibfnamefont
  {M.}~\bibnamefont {Greven}}, \bibinfo {author} {\bibfnamefont
  {S.}~\bibnamefont {Lupi}}, \bibinfo {author} {\bibfnamefont {A.}~\bibnamefont
  {Damascelli}}, \bibinfo {author} {\bibfnamefont {D.}~\bibnamefont {Brida}},
  \bibinfo {author} {\bibfnamefont {M.}~\bibnamefont {Capone}}, \bibinfo
  {author} {\bibfnamefont {J.}~\bibnamefont {Bon{\v{c}}a}}, \bibinfo {author}
  {\bibfnamefont {G.}~\bibnamefont {Cerullo}},\ and\ \bibinfo {author}
  {\bibfnamefont {C.}~\bibnamefont {Giannetti}},\ }\bibfield  {title} {\bibinfo
  {title} {Snapshots of the retarded interaction of charge carriers with
  ultrafast fluctuations in cuprates},\ }\href
  {https://doi.org/10.1038/nphys3265} {\bibfield  {journal} {\bibinfo
  {journal} {Nature Physics}\ }\textbf {\bibinfo {volume} {11}},\ \bibinfo
  {pages} {421} (\bibinfo {year} {2015})}\BibitemShut {NoStop}%
\bibitem [{\citenamefont {Giannetti}\ \emph {et~al.}(2016)\citenamefont
  {Giannetti}, \citenamefont {Capone}, \citenamefont {Fausti}, \citenamefont
  {Fabrizio}, \citenamefont {Parmigiani},\ and\ \citenamefont
  {Mihailovic}}]{doi:10.1080/00018732.2016.1194044}%
  \BibitemOpen
  \bibfield  {author} {\bibinfo {author} {\bibfnamefont {C.}~\bibnamefont
  {Giannetti}}, \bibinfo {author} {\bibfnamefont {M.}~\bibnamefont {Capone}},
  \bibinfo {author} {\bibfnamefont {D.}~\bibnamefont {Fausti}}, \bibinfo
  {author} {\bibfnamefont {M.}~\bibnamefont {Fabrizio}}, \bibinfo {author}
  {\bibfnamefont {F.}~\bibnamefont {Parmigiani}},\ and\ \bibinfo {author}
  {\bibfnamefont {D.}~\bibnamefont {Mihailovic}},\ }\bibfield  {title}
  {\bibinfo {title} {Ultrafast optical spectroscopy of strongly correlated
  materials and high-temperature superconductors: a non-equilibrium approach},\
  }\href {https://doi.org/10.1080/00018732.2016.1194044} {\bibfield  {journal}
  {\bibinfo  {journal} {Advances in Physics}\ }\textbf {\bibinfo {volume}
  {65}},\ \bibinfo {pages} {58} (\bibinfo {year} {2016})}\BibitemShut {NoStop}%
\bibitem [{\citenamefont {Basov}\ \emph {et~al.}(2011)\citenamefont {Basov},
  \citenamefont {Averitt}, \citenamefont {van~der Marel}, \citenamefont
  {Dressel},\ and\ \citenamefont {Haule}}]{RevModPhys.83.471}%
  \BibitemOpen
  \bibfield  {author} {\bibinfo {author} {\bibfnamefont {D.~N.}\ \bibnamefont
  {Basov}}, \bibinfo {author} {\bibfnamefont {R.~D.}\ \bibnamefont {Averitt}},
  \bibinfo {author} {\bibfnamefont {D.}~\bibnamefont {van~der Marel}}, \bibinfo
  {author} {\bibfnamefont {M.}~\bibnamefont {Dressel}},\ and\ \bibinfo {author}
  {\bibfnamefont {K.}~\bibnamefont {Haule}},\ }\bibfield  {title} {\bibinfo
  {title} {Electrodynamics of correlated electron materials},\ }\href
  {https://doi.org/10.1103/RevModPhys.83.471} {\bibfield  {journal} {\bibinfo
  {journal} {Rev. Mod. Phys.}\ }\textbf {\bibinfo {volume} {83}},\ \bibinfo
  {pages} {471} (\bibinfo {year} {2011})}\BibitemShut {NoStop}%
\bibitem [{\citenamefont {Orenstein}(2012)}]{doi:10.1063/PT.3.1717}%
  \BibitemOpen
  \bibfield  {author} {\bibinfo {author} {\bibfnamefont {J.}~\bibnamefont
  {Orenstein}},\ }\bibfield  {title} {\bibinfo {title} {Ultrafast spectroscopy
  of quantum materials},\ }\href {https://doi.org/10.1063/PT.3.1717} {\bibfield
   {journal} {\bibinfo  {journal} {Physics Today}\ }\textbf {\bibinfo {volume}
  {65}},\ \bibinfo {pages} {44} (\bibinfo {year} {2012})}\BibitemShut {NoStop}%
\bibitem [{\citenamefont {Basov}\ \emph {et~al.}(2017)\citenamefont {Basov},
  \citenamefont {Averitt},\ and\ \citenamefont {Hsieh}}]{Basov2017}%
  \BibitemOpen
  \bibfield  {author} {\bibinfo {author} {\bibfnamefont {D.~N.}\ \bibnamefont
  {Basov}}, \bibinfo {author} {\bibfnamefont {R.~D.}\ \bibnamefont {Averitt}},\
  and\ \bibinfo {author} {\bibfnamefont {D.}~\bibnamefont {Hsieh}},\ }\bibfield
   {title} {\bibinfo {title} {Towards properties on demand in quantum
  materials},\ }\href {https://doi.org/10.1038/nmat5017} {\bibfield  {journal}
  {\bibinfo  {journal} {Nature Materials}\ }\textbf {\bibinfo {volume} {16}},\
  \bibinfo {pages} {1077} (\bibinfo {year} {2017})}\BibitemShut {NoStop}%
\bibitem [{\citenamefont {de~la Torre}\ \emph {et~al.}(2021)\citenamefont
  {de~la Torre}, \citenamefont {Kennes}, \citenamefont {Claassen},
  \citenamefont {Gerber}, \citenamefont {McIver},\ and\ \citenamefont
  {Sentef}}]{RevModPhys.93.041002}%
  \BibitemOpen
  \bibfield  {author} {\bibinfo {author} {\bibfnamefont {A.}~\bibnamefont
  {de~la Torre}}, \bibinfo {author} {\bibfnamefont {D.~M.}\ \bibnamefont
  {Kennes}}, \bibinfo {author} {\bibfnamefont {M.}~\bibnamefont {Claassen}},
  \bibinfo {author} {\bibfnamefont {S.}~\bibnamefont {Gerber}}, \bibinfo
  {author} {\bibfnamefont {J.~W.}\ \bibnamefont {McIver}},\ and\ \bibinfo
  {author} {\bibfnamefont {M.~A.}\ \bibnamefont {Sentef}},\ }\bibfield  {title}
  {\bibinfo {title} {Colloquium: Nonthermal pathways to ultrafast control in
  quantum materials},\ }\href {https://doi.org/10.1103/RevModPhys.93.041002}
  {\bibfield  {journal} {\bibinfo  {journal} {Rev. Mod. Phys.}\ }\textbf
  {\bibinfo {volume} {93}},\ \bibinfo {pages} {041002} (\bibinfo {year}
  {2021})}\BibitemShut {NoStop}%
\bibitem [{\citenamefont {Carusotto}\ and\ \citenamefont
  {Ciuti}(2013)}]{RevModPhys.85.299}%
  \BibitemOpen
  \bibfield  {author} {\bibinfo {author} {\bibfnamefont {I.}~\bibnamefont
  {Carusotto}}\ and\ \bibinfo {author} {\bibfnamefont {C.}~\bibnamefont
  {Ciuti}},\ }\bibfield  {title} {\bibinfo {title} {Quantum fluids of light},\
  }\href {https://doi.org/10.1103/RevModPhys.85.299} {\bibfield  {journal}
  {\bibinfo  {journal} {Rev. Mod. Phys.}\ }\textbf {\bibinfo {volume} {85}},\
  \bibinfo {pages} {299} (\bibinfo {year} {2013})}\BibitemShut {NoStop}%
\bibitem [{\citenamefont {Chiacchio}\ and\ \citenamefont
  {Nunnenkamp}(2018)}]{PhysRevA.98.023617}%
  \BibitemOpen
  \bibfield  {author} {\bibinfo {author} {\bibfnamefont {E.~I.~R.}\
  \bibnamefont {Chiacchio}}\ and\ \bibinfo {author} {\bibfnamefont
  {A.}~\bibnamefont {Nunnenkamp}},\ }\bibfield  {title} {\bibinfo {title}
  {Emergence of continuous rotational symmetries in ultracold atoms coupled to
  optical cavities},\ }\href {https://doi.org/10.1103/PhysRevA.98.023617}
  {\bibfield  {journal} {\bibinfo  {journal} {Phys. Rev. A}\ }\textbf {\bibinfo
  {volume} {98}},\ \bibinfo {pages} {023617} (\bibinfo {year}
  {2018})}\BibitemShut {NoStop}%
\bibitem [{\citenamefont {Lambert}\ \emph {et~al.}(2019)\citenamefont
  {Lambert}, \citenamefont {Ahmed}, \citenamefont {Cirio},\ and\ \citenamefont
  {Nori}}]{Lambert2019}%
  \BibitemOpen
  \bibfield  {author} {\bibinfo {author} {\bibfnamefont {N.}~\bibnamefont
  {Lambert}}, \bibinfo {author} {\bibfnamefont {S.}~\bibnamefont {Ahmed}},
  \bibinfo {author} {\bibfnamefont {M.}~\bibnamefont {Cirio}},\ and\ \bibinfo
  {author} {\bibfnamefont {F.}~\bibnamefont {Nori}},\ }\bibfield  {title}
  {\bibinfo {title} {Modelling the ultra-strongly coupled spin-boson model with
  unphysical modes},\ }\href {https://doi.org/10.1038/s41467-019-11656-1}
  {\bibfield  {journal} {\bibinfo  {journal} {Nature Communications}\ }\textbf
  {\bibinfo {volume} {10}},\ \bibinfo {pages} {3721} (\bibinfo {year}
  {2019})}\BibitemShut {NoStop}%
\bibitem [{\citenamefont {Gr{\"a}tzel}(2001)}]{Gratzel2001}%
  \BibitemOpen
  \bibfield  {author} {\bibinfo {author} {\bibfnamefont {M.}~\bibnamefont
  {Gr{\"a}tzel}},\ }\bibfield  {title} {\bibinfo {title} {Photoelectrochemical
  cells},\ }\href {https://doi.org/10.1038/35104607} {\bibfield  {journal}
  {\bibinfo  {journal} {Nature}\ }\textbf {\bibinfo {volume} {414}},\ \bibinfo
  {pages} {338} (\bibinfo {year} {2001})}\BibitemShut {NoStop}%
\bibitem [{\citenamefont {Wright}\ \emph {et~al.}(2016)\citenamefont {Wright},
  \citenamefont {Verdi}, \citenamefont {Milot}, \citenamefont {Eperon},
  \citenamefont {P{\'e}rez-Osorio}, \citenamefont {Snaith}, \citenamefont
  {Giustino}, \citenamefont {Johnston},\ and\ \citenamefont
  {Herz}}]{Wright2016}%
  \BibitemOpen
  \bibfield  {author} {\bibinfo {author} {\bibfnamefont {A.~D.}\ \bibnamefont
  {Wright}}, \bibinfo {author} {\bibfnamefont {C.}~\bibnamefont {Verdi}},
  \bibinfo {author} {\bibfnamefont {R.~L.}\ \bibnamefont {Milot}}, \bibinfo
  {author} {\bibfnamefont {G.~E.}\ \bibnamefont {Eperon}}, \bibinfo {author}
  {\bibfnamefont {M.~A.}\ \bibnamefont {P{\'e}rez-Osorio}}, \bibinfo {author}
  {\bibfnamefont {H.~J.}\ \bibnamefont {Snaith}}, \bibinfo {author}
  {\bibfnamefont {F.}~\bibnamefont {Giustino}}, \bibinfo {author}
  {\bibfnamefont {M.~B.}\ \bibnamefont {Johnston}},\ and\ \bibinfo {author}
  {\bibfnamefont {L.~M.}\ \bibnamefont {Herz}},\ }\bibfield  {title} {\bibinfo
  {title} {Electron--phonon coupling in hybrid lead halide perovskites},\
  }\href {https://doi.org/10.1038/ncomms11755} {\bibfield  {journal} {\bibinfo
  {journal} {Nature Communications}\ }\textbf {\bibinfo {volume} {7}},\
  \bibinfo {pages} {11755} (\bibinfo {year} {2016})}\BibitemShut {NoStop}%
\bibitem [{\citenamefont {Rini}\ \emph {et~al.}(2007)\citenamefont {Rini},
  \citenamefont {Tobey}, \citenamefont {Dean}, \citenamefont {Itatani},
  \citenamefont {Tomioka}, \citenamefont {Tokura}, \citenamefont {Schoenlein},\
  and\ \citenamefont {Cavalleri}}]{Rini2007}%
  \BibitemOpen
  \bibfield  {author} {\bibinfo {author} {\bibfnamefont {M.}~\bibnamefont
  {Rini}}, \bibinfo {author} {\bibfnamefont {R.}~\bibnamefont {Tobey}},
  \bibinfo {author} {\bibfnamefont {N.}~\bibnamefont {Dean}}, \bibinfo {author}
  {\bibfnamefont {J.}~\bibnamefont {Itatani}}, \bibinfo {author} {\bibfnamefont
  {Y.}~\bibnamefont {Tomioka}}, \bibinfo {author} {\bibfnamefont
  {Y.}~\bibnamefont {Tokura}}, \bibinfo {author} {\bibfnamefont {R.~W.}\
  \bibnamefont {Schoenlein}},\ and\ \bibinfo {author} {\bibfnamefont
  {A.}~\bibnamefont {Cavalleri}},\ }\bibfield  {title} {\bibinfo {title}
  {Control of the electronic phase of a manganite by mode-selective vibrational
  excitation},\ }\href {https://doi.org/10.1038/nature06119} {\bibfield
  {journal} {\bibinfo  {journal} {Nature}\ }\textbf {\bibinfo {volume} {449}},\
  \bibinfo {pages} {72} (\bibinfo {year} {2007})}\BibitemShut {NoStop}%
\bibitem [{\citenamefont {Mitrano}\ \emph {et~al.}(2016)\citenamefont
  {Mitrano}, \citenamefont {Cantaluppi}, \citenamefont {Nicoletti},
  \citenamefont {Kaiser}, \citenamefont {Perucchi}, \citenamefont {Lupi},
  \citenamefont {Di~Pietro}, \citenamefont {Pontiroli}, \citenamefont
  {Ricc{\`o}}, \citenamefont {Clark}, \citenamefont {Jaksch},\ and\
  \citenamefont {Cavalleri}}]{Mitrano2016}%
  \BibitemOpen
  \bibfield  {author} {\bibinfo {author} {\bibfnamefont {M.}~\bibnamefont
  {Mitrano}}, \bibinfo {author} {\bibfnamefont {A.}~\bibnamefont {Cantaluppi}},
  \bibinfo {author} {\bibfnamefont {D.}~\bibnamefont {Nicoletti}}, \bibinfo
  {author} {\bibfnamefont {S.}~\bibnamefont {Kaiser}}, \bibinfo {author}
  {\bibfnamefont {A.}~\bibnamefont {Perucchi}}, \bibinfo {author}
  {\bibfnamefont {S.}~\bibnamefont {Lupi}}, \bibinfo {author} {\bibfnamefont
  {P.}~\bibnamefont {Di~Pietro}}, \bibinfo {author} {\bibfnamefont
  {D.}~\bibnamefont {Pontiroli}}, \bibinfo {author} {\bibfnamefont
  {M.}~\bibnamefont {Ricc{\`o}}}, \bibinfo {author} {\bibfnamefont {S.~R.}\
  \bibnamefont {Clark}}, \bibinfo {author} {\bibfnamefont {D.}~\bibnamefont
  {Jaksch}},\ and\ \bibinfo {author} {\bibfnamefont {A.}~\bibnamefont
  {Cavalleri}},\ }\bibfield  {title} {\bibinfo {title} {Possible light-induced
  superconductivity in {K}$_{3}${C}$_{60}$ at high temperature},\ }\href
  {https://doi.org/10.1038/nature16522} {\bibfield  {journal} {\bibinfo
  {journal} {Nature}\ }\textbf {\bibinfo {volume} {530}},\ \bibinfo {pages}
  {461} (\bibinfo {year} {2016})}\BibitemShut {NoStop}%
\bibitem [{\citenamefont {Malinauskas}\ \emph {et~al.}(2016)\citenamefont
  {Malinauskas}, \citenamefont {{\v{Z}}ukauskas}, \citenamefont {Hasegawa},
  \citenamefont {Hayasaki}, \citenamefont {Mizeikis}, \citenamefont
  {Buividas},\ and\ \citenamefont {Juodkazis}}]{Malinauskas2016}%
  \BibitemOpen
  \bibfield  {author} {\bibinfo {author} {\bibfnamefont {M.}~\bibnamefont
  {Malinauskas}}, \bibinfo {author} {\bibfnamefont {A.}~\bibnamefont
  {{\v{Z}}ukauskas}}, \bibinfo {author} {\bibfnamefont {S.}~\bibnamefont
  {Hasegawa}}, \bibinfo {author} {\bibfnamefont {Y.}~\bibnamefont {Hayasaki}},
  \bibinfo {author} {\bibfnamefont {V.}~\bibnamefont {Mizeikis}}, \bibinfo
  {author} {\bibfnamefont {R.}~\bibnamefont {Buividas}},\ and\ \bibinfo
  {author} {\bibfnamefont {S.}~\bibnamefont {Juodkazis}},\ }\bibfield  {title}
  {\bibinfo {title} {Ultrafast laser processing of materials: from science to
  industry},\ }\href {https://doi.org/10.1038/lsa.2016.133} {\bibfield
  {journal} {\bibinfo  {journal} {Light: Science {\&} Applications}\ }\textbf
  {\bibinfo {volume} {5}},\ \bibinfo {pages} {e16133} (\bibinfo {year}
  {2016})}\BibitemShut {NoStop}%
\bibitem [{\citenamefont {Mankowsky}\ \emph {et~al.}(2016)\citenamefont
  {Mankowsky}, \citenamefont {Först},\ and\ \citenamefont
  {Cavalleri}}]{Mankowsky_2016}%
  \BibitemOpen
  \bibfield  {author} {\bibinfo {author} {\bibfnamefont {R.}~\bibnamefont
  {Mankowsky}}, \bibinfo {author} {\bibfnamefont {M.}~\bibnamefont {Först}},\
  and\ \bibinfo {author} {\bibfnamefont {A.}~\bibnamefont {Cavalleri}},\
  }\bibfield  {title} {\bibinfo {title} {Non-equilibrium control of complex
  solids by nonlinear phononics},\ }\href
  {https://doi.org/10.1088/0034-4885/79/6/064503} {\bibfield  {journal}
  {\bibinfo  {journal} {Reports on Progress in Physics}\ }\textbf {\bibinfo
  {volume} {79}},\ \bibinfo {pages} {064503} (\bibinfo {year}
  {2016})}\BibitemShut {NoStop}%
\bibitem [{\citenamefont {McQueeney}\ \emph {et~al.}(2001)\citenamefont
  {McQueeney}, \citenamefont {Sarrao}, \citenamefont {Pagliuso}, \citenamefont
  {Stephens},\ and\ \citenamefont {Osborn}}]{PhysRevLett.87.077001}%
  \BibitemOpen
  \bibfield  {author} {\bibinfo {author} {\bibfnamefont {R.~J.}\ \bibnamefont
  {McQueeney}}, \bibinfo {author} {\bibfnamefont {J.~L.}\ \bibnamefont
  {Sarrao}}, \bibinfo {author} {\bibfnamefont {P.~G.}\ \bibnamefont
  {Pagliuso}}, \bibinfo {author} {\bibfnamefont {P.~W.}\ \bibnamefont
  {Stephens}},\ and\ \bibinfo {author} {\bibfnamefont {R.}~\bibnamefont
  {Osborn}},\ }\bibfield  {title} {\bibinfo {title} {Mixed lattice and
  electronic states in high-temperature superconductors},\ }\href
  {https://doi.org/10.1103/PhysRevLett.87.077001} {\bibfield  {journal}
  {\bibinfo  {journal} {Phys. Rev. Lett.}\ }\textbf {\bibinfo {volume} {87}},\
  \bibinfo {pages} {077001} (\bibinfo {year} {2001})}\BibitemShut {NoStop}%
\bibitem [{\citenamefont {Tajima}\ \emph {et~al.}(2005)\citenamefont {Tajima},
  \citenamefont {Fudamoto}, \citenamefont {Kakeshita}, \citenamefont
  {Gorshunov}, \citenamefont {\ifmmode~\check{Z}\else \v{Z}\fi{}elezn\'y},
  \citenamefont {Kojima}, \citenamefont {Dressel},\ and\ \citenamefont
  {Uchida}}]{PhysRevB.71.094508}%
  \BibitemOpen
  \bibfield  {author} {\bibinfo {author} {\bibfnamefont {S.}~\bibnamefont
  {Tajima}}, \bibinfo {author} {\bibfnamefont {Y.}~\bibnamefont {Fudamoto}},
  \bibinfo {author} {\bibfnamefont {T.}~\bibnamefont {Kakeshita}}, \bibinfo
  {author} {\bibfnamefont {B.}~\bibnamefont {Gorshunov}}, \bibinfo {author}
  {\bibfnamefont {V.}~\bibnamefont {\ifmmode~\check{Z}\else
  \v{Z}\fi{}elezn\'y}}, \bibinfo {author} {\bibfnamefont {K.~M.}\ \bibnamefont
  {Kojima}}, \bibinfo {author} {\bibfnamefont {M.}~\bibnamefont {Dressel}},\
  and\ \bibinfo {author} {\bibfnamefont {S.}~\bibnamefont {Uchida}},\
  }\bibfield  {title} {\bibinfo {title} {In-plane optical conductivity of
  {La}$_{2-x}${Sr}$_x${Cu}{O}$_4$: Reduced superconducting condensate and
  residual {D}rude-like response},\ }\href
  {https://doi.org/10.1103/PhysRevB.71.094508} {\bibfield  {journal} {\bibinfo
  {journal} {Phys. Rev. B}\ }\textbf {\bibinfo {volume} {71}},\ \bibinfo
  {pages} {094508} (\bibinfo {year} {2005})}\BibitemShut {NoStop}%
\bibitem [{\citenamefont {Perfetti}\ \emph {et~al.}(2006)\citenamefont
  {Perfetti}, \citenamefont {Loukakos}, \citenamefont {Lisowski}, \citenamefont
  {Bovensiepen}, \citenamefont {Berger}, \citenamefont {Biermann},
  \citenamefont {Cornaglia}, \citenamefont {Georges},\ and\ \citenamefont
  {Wolf}}]{PhysRevLett.97.067402}%
  \BibitemOpen
  \bibfield  {author} {\bibinfo {author} {\bibfnamefont {L.}~\bibnamefont
  {Perfetti}}, \bibinfo {author} {\bibfnamefont {P.~A.}\ \bibnamefont
  {Loukakos}}, \bibinfo {author} {\bibfnamefont {M.}~\bibnamefont {Lisowski}},
  \bibinfo {author} {\bibfnamefont {U.}~\bibnamefont {Bovensiepen}}, \bibinfo
  {author} {\bibfnamefont {H.}~\bibnamefont {Berger}}, \bibinfo {author}
  {\bibfnamefont {S.}~\bibnamefont {Biermann}}, \bibinfo {author}
  {\bibfnamefont {P.~S.}\ \bibnamefont {Cornaglia}}, \bibinfo {author}
  {\bibfnamefont {A.}~\bibnamefont {Georges}},\ and\ \bibinfo {author}
  {\bibfnamefont {M.}~\bibnamefont {Wolf}},\ }\bibfield  {title} {\bibinfo
  {title} {Time evolution of the electronic structure of
  1\textit{{T}}-{TaS}$_2$ through the insulator-metal transition},\ }\href
  {https://doi.org/10.1103/PhysRevLett.97.067402} {\bibfield  {journal}
  {\bibinfo  {journal} {Phys. Rev. Lett.}\ }\textbf {\bibinfo {volume} {97}},\
  \bibinfo {pages} {067402} (\bibinfo {year} {2006})}\BibitemShut {NoStop}%
\bibitem [{\citenamefont {Hellmann}\ \emph {et~al.}(2010)\citenamefont
  {Hellmann}, \citenamefont {Beye}, \citenamefont {Sohrt}, \citenamefont
  {Rohwer}, \citenamefont {Sorgenfrei}, \citenamefont {Redlin}, \citenamefont
  {Kall\"ane}, \citenamefont {Marczynski-B\"uhlow}, \citenamefont {Hennies},
  \citenamefont {Bauer}, \citenamefont {F\"ohlisch}, \citenamefont {Kipp},
  \citenamefont {Wurth},\ and\ \citenamefont
  {Rossnagel}}]{PhysRevLett.105.187401}%
  \BibitemOpen
  \bibfield  {author} {\bibinfo {author} {\bibfnamefont {S.}~\bibnamefont
  {Hellmann}}, \bibinfo {author} {\bibfnamefont {M.}~\bibnamefont {Beye}},
  \bibinfo {author} {\bibfnamefont {C.}~\bibnamefont {Sohrt}}, \bibinfo
  {author} {\bibfnamefont {T.}~\bibnamefont {Rohwer}}, \bibinfo {author}
  {\bibfnamefont {F.}~\bibnamefont {Sorgenfrei}}, \bibinfo {author}
  {\bibfnamefont {H.}~\bibnamefont {Redlin}}, \bibinfo {author} {\bibfnamefont
  {M.}~\bibnamefont {Kall\"ane}}, \bibinfo {author} {\bibfnamefont
  {M.}~\bibnamefont {Marczynski-B\"uhlow}}, \bibinfo {author} {\bibfnamefont
  {F.}~\bibnamefont {Hennies}}, \bibinfo {author} {\bibfnamefont
  {M.}~\bibnamefont {Bauer}}, \bibinfo {author} {\bibfnamefont
  {A.}~\bibnamefont {F\"ohlisch}}, \bibinfo {author} {\bibfnamefont
  {L.}~\bibnamefont {Kipp}}, \bibinfo {author} {\bibfnamefont {W.}~\bibnamefont
  {Wurth}},\ and\ \bibinfo {author} {\bibfnamefont {K.}~\bibnamefont
  {Rossnagel}},\ }\bibfield  {title} {\bibinfo {title} {Ultrafast melting of a
  charge-density wave in the mott insulator 1\textit{{T}}-{TaS}$_2$},\ }\href
  {https://doi.org/10.1103/PhysRevLett.105.187401} {\bibfield  {journal}
  {\bibinfo  {journal} {Phys. Rev. Lett.}\ }\textbf {\bibinfo {volume} {105}},\
  \bibinfo {pages} {187401} (\bibinfo {year} {2010})}\BibitemShut {NoStop}%
\bibitem [{\citenamefont {Hellmann}\ \emph {et~al.}(2012)\citenamefont
  {Hellmann}, \citenamefont {Rohwer}, \citenamefont {Kall{\"a}ne},
  \citenamefont {Hanff}, \citenamefont {Sohrt}, \citenamefont {Stange},
  \citenamefont {Carr}, \citenamefont {Murnane}, \citenamefont {Kapteyn},
  \citenamefont {Kipp}, \citenamefont {Bauer},\ and\ \citenamefont
  {Rossnagel}}]{Hellmann2012}%
  \BibitemOpen
  \bibfield  {author} {\bibinfo {author} {\bibfnamefont {S.}~\bibnamefont
  {Hellmann}}, \bibinfo {author} {\bibfnamefont {T.}~\bibnamefont {Rohwer}},
  \bibinfo {author} {\bibfnamefont {M.}~\bibnamefont {Kall{\"a}ne}}, \bibinfo
  {author} {\bibfnamefont {K.}~\bibnamefont {Hanff}}, \bibinfo {author}
  {\bibfnamefont {C.}~\bibnamefont {Sohrt}}, \bibinfo {author} {\bibfnamefont
  {A.}~\bibnamefont {Stange}}, \bibinfo {author} {\bibfnamefont
  {A.}~\bibnamefont {Carr}}, \bibinfo {author} {\bibfnamefont {M.~M.}\
  \bibnamefont {Murnane}}, \bibinfo {author} {\bibfnamefont {H.~C.}\
  \bibnamefont {Kapteyn}}, \bibinfo {author} {\bibfnamefont {L.}~\bibnamefont
  {Kipp}}, \bibinfo {author} {\bibfnamefont {M.}~\bibnamefont {Bauer}},\ and\
  \bibinfo {author} {\bibfnamefont {K.}~\bibnamefont {Rossnagel}},\ }\bibfield
  {title} {\bibinfo {title} {Time-domain classification of charge-density-wave
  insulators},\ }\href {https://doi.org/10.1038/ncomms2078} {\bibfield
  {journal} {\bibinfo  {journal} {Nature Communications}\ }\textbf {\bibinfo
  {volume} {3}},\ \bibinfo {pages} {1069} (\bibinfo {year} {2012})}\BibitemShut
  {NoStop}%
\bibitem [{\citenamefont {Rohwer}\ \emph {et~al.}(2011)\citenamefont {Rohwer},
  \citenamefont {Hellmann}, \citenamefont {Wiesenmayer}, \citenamefont {Sohrt},
  \citenamefont {Stange}, \citenamefont {Slomski}, \citenamefont {Carr},
  \citenamefont {Liu}, \citenamefont {Avila}, \citenamefont {Kall{\"a}ne},
  \citenamefont {Mathias}, \citenamefont {Kipp}, \citenamefont {Rossnagel},\
  and\ \citenamefont {Bauer}}]{Rohwer2011}%
  \BibitemOpen
  \bibfield  {author} {\bibinfo {author} {\bibfnamefont {T.}~\bibnamefont
  {Rohwer}}, \bibinfo {author} {\bibfnamefont {S.}~\bibnamefont {Hellmann}},
  \bibinfo {author} {\bibfnamefont {M.}~\bibnamefont {Wiesenmayer}}, \bibinfo
  {author} {\bibfnamefont {C.}~\bibnamefont {Sohrt}}, \bibinfo {author}
  {\bibfnamefont {A.}~\bibnamefont {Stange}}, \bibinfo {author} {\bibfnamefont
  {B.}~\bibnamefont {Slomski}}, \bibinfo {author} {\bibfnamefont
  {A.}~\bibnamefont {Carr}}, \bibinfo {author} {\bibfnamefont {Y.}~\bibnamefont
  {Liu}}, \bibinfo {author} {\bibfnamefont {L.~M.}\ \bibnamefont {Avila}},
  \bibinfo {author} {\bibfnamefont {M.}~\bibnamefont {Kall{\"a}ne}}, \bibinfo
  {author} {\bibfnamefont {S.}~\bibnamefont {Mathias}}, \bibinfo {author}
  {\bibfnamefont {L.}~\bibnamefont {Kipp}}, \bibinfo {author} {\bibfnamefont
  {K.}~\bibnamefont {Rossnagel}},\ and\ \bibinfo {author} {\bibfnamefont
  {M.}~\bibnamefont {Bauer}},\ }\bibfield  {title} {\bibinfo {title} {Collapse
  of long-range charge order tracked by time-resolved photoemission at high
  momenta},\ }\href {https://doi.org/10.1038/nature09829} {\bibfield  {journal}
  {\bibinfo  {journal} {Nature}\ }\textbf {\bibinfo {volume} {471}},\ \bibinfo
  {pages} {490} (\bibinfo {year} {2011})}\BibitemShut {NoStop}%
\bibitem [{\citenamefont {Chan}\ \emph {et~al.}(2015)\citenamefont {Chan},
  \citenamefont {Lee},\ and\ \citenamefont
  {Gopalakrishnan}}]{PhysRevA.91.051601}%
  \BibitemOpen
  \bibfield  {author} {\bibinfo {author} {\bibfnamefont {C.-K.}\ \bibnamefont
  {Chan}}, \bibinfo {author} {\bibfnamefont {T.~E.}\ \bibnamefont {Lee}},\ and\
  \bibinfo {author} {\bibfnamefont {S.}~\bibnamefont {Gopalakrishnan}},\
  }\bibfield  {title} {\bibinfo {title} {Limit-cycle phase in
  driven-dissipative spin systems},\ }\href
  {https://doi.org/10.1103/PhysRevA.91.051601} {\bibfield  {journal} {\bibinfo
  {journal} {Phys. Rev. A}\ }\textbf {\bibinfo {volume} {91}},\ \bibinfo
  {pages} {051601} (\bibinfo {year} {2015})}\BibitemShut {NoStop}%
\bibitem [{\citenamefont {Wilson}\ \emph {et~al.}(2016)\citenamefont {Wilson},
  \citenamefont {Mahmud}, \citenamefont {Hu}, \citenamefont {Gorshkov},
  \citenamefont {Hafezi},\ and\ \citenamefont
  {Foss-Feig}}]{PhysRevA.94.033801}%
  \BibitemOpen
  \bibfield  {author} {\bibinfo {author} {\bibfnamefont {R.~M.}\ \bibnamefont
  {Wilson}}, \bibinfo {author} {\bibfnamefont {K.~W.}\ \bibnamefont {Mahmud}},
  \bibinfo {author} {\bibfnamefont {A.}~\bibnamefont {Hu}}, \bibinfo {author}
  {\bibfnamefont {A.~V.}\ \bibnamefont {Gorshkov}}, \bibinfo {author}
  {\bibfnamefont {M.}~\bibnamefont {Hafezi}},\ and\ \bibinfo {author}
  {\bibfnamefont {M.}~\bibnamefont {Foss-Feig}},\ }\bibfield  {title} {\bibinfo
  {title} {Collective phases of strongly interacting cavity photons},\ }\href
  {https://doi.org/10.1103/PhysRevA.94.033801} {\bibfield  {journal} {\bibinfo
  {journal} {Phys. Rev. A}\ }\textbf {\bibinfo {volume} {94}},\ \bibinfo
  {pages} {033801} (\bibinfo {year} {2016})}\BibitemShut {NoStop}%
\bibitem [{\citenamefont {Yarmohammadi}\ \emph {et~al.}(2021)\citenamefont
  {Yarmohammadi}, \citenamefont {Meyer}, \citenamefont {Fauseweh},
  \citenamefont {Normand},\ and\ \citenamefont
  {Uhrig}}]{yarmohammadi2020dynamical}%
  \BibitemOpen
  \bibfield  {author} {\bibinfo {author} {\bibfnamefont {M.}~\bibnamefont
  {Yarmohammadi}}, \bibinfo {author} {\bibfnamefont {C.}~\bibnamefont {Meyer}},
  \bibinfo {author} {\bibfnamefont {B.}~\bibnamefont {Fauseweh}}, \bibinfo
  {author} {\bibfnamefont {B.}~\bibnamefont {Normand}},\ and\ \bibinfo {author}
  {\bibfnamefont {G.~S.}\ \bibnamefont {Uhrig}},\ }\bibfield  {title} {\bibinfo
  {title} {Dynamical properties of a driven dissipative dimerized
  ${S}=\frac{1}{2}$ chain},\ }\href
  {https://doi.org/10.1103/PhysRevB.103.045132} {\bibfield  {journal} {\bibinfo
   {journal} {Phys. Rev. B}\ }\textbf {\bibinfo {volume} {103}},\ \bibinfo
  {pages} {045132} (\bibinfo {year} {2021})}\BibitemShut {NoStop}%
\bibitem [{\citenamefont {Schir\'o}\ \emph {et~al.}(2016)\citenamefont
  {Schir\'o}, \citenamefont {Joshi}, \citenamefont {Bordyuh}, \citenamefont
  {Fazio}, \citenamefont {Keeling},\ and\ \citenamefont
  {T\"ureci}}]{PhysRevLett.116.143603}%
  \BibitemOpen
  \bibfield  {author} {\bibinfo {author} {\bibfnamefont {M.}~\bibnamefont
  {Schir\'o}}, \bibinfo {author} {\bibfnamefont {C.}~\bibnamefont {Joshi}},
  \bibinfo {author} {\bibfnamefont {M.}~\bibnamefont {Bordyuh}}, \bibinfo
  {author} {\bibfnamefont {R.}~\bibnamefont {Fazio}}, \bibinfo {author}
  {\bibfnamefont {J.}~\bibnamefont {Keeling}},\ and\ \bibinfo {author}
  {\bibfnamefont {H.~E.}\ \bibnamefont {T\"ureci}},\ }\bibfield  {title}
  {\bibinfo {title} {Exotic attractors of the nonequilibrium {R}abi-{H}ubbard
  model},\ }\href {https://doi.org/10.1103/PhysRevLett.116.143603} {\bibfield
  {journal} {\bibinfo  {journal} {Phys. Rev. Lett.}\ }\textbf {\bibinfo
  {volume} {116}},\ \bibinfo {pages} {143603} (\bibinfo {year}
  {2016})}\BibitemShut {NoStop}%
\bibitem [{\citenamefont {Jin}\ \emph {et~al.}(2013)\citenamefont {Jin},
  \citenamefont {Rossini}, \citenamefont {Fazio}, \citenamefont {Leib},\ and\
  \citenamefont {Hartmann}}]{PhysRevLett.110.163605}%
  \BibitemOpen
  \bibfield  {author} {\bibinfo {author} {\bibfnamefont {J.}~\bibnamefont
  {Jin}}, \bibinfo {author} {\bibfnamefont {D.}~\bibnamefont {Rossini}},
  \bibinfo {author} {\bibfnamefont {R.}~\bibnamefont {Fazio}}, \bibinfo
  {author} {\bibfnamefont {M.}~\bibnamefont {Leib}},\ and\ \bibinfo {author}
  {\bibfnamefont {M.~J.}\ \bibnamefont {Hartmann}},\ }\bibfield  {title}
  {\bibinfo {title} {Photon solid phases in driven arrays of nonlinearly
  coupled cavities},\ }\href {https://doi.org/10.1103/PhysRevLett.110.163605}
  {\bibfield  {journal} {\bibinfo  {journal} {Phys. Rev. Lett.}\ }\textbf
  {\bibinfo {volume} {110}},\ \bibinfo {pages} {163605} (\bibinfo {year}
  {2013})}\BibitemShut {NoStop}%
\bibitem [{\citenamefont {Jin}\ \emph {et~al.}(2014)\citenamefont {Jin},
  \citenamefont {Rossini}, \citenamefont {Leib}, \citenamefont {Hartmann},\
  and\ \citenamefont {Fazio}}]{PhysRevA.90.023827}%
  \BibitemOpen
  \bibfield  {author} {\bibinfo {author} {\bibfnamefont {J.}~\bibnamefont
  {Jin}}, \bibinfo {author} {\bibfnamefont {D.}~\bibnamefont {Rossini}},
  \bibinfo {author} {\bibfnamefont {M.}~\bibnamefont {Leib}}, \bibinfo {author}
  {\bibfnamefont {M.~J.}\ \bibnamefont {Hartmann}},\ and\ \bibinfo {author}
  {\bibfnamefont {R.}~\bibnamefont {Fazio}},\ }\bibfield  {title} {\bibinfo
  {title} {Steady-state phase diagram of a driven qed-cavity array with
  cross-{K}err nonlinearities},\ }\href
  {https://doi.org/10.1103/PhysRevA.90.023827} {\bibfield  {journal} {\bibinfo
  {journal} {Phys. Rev. A}\ }\textbf {\bibinfo {volume} {90}},\ \bibinfo
  {pages} {023827} (\bibinfo {year} {2014})}\BibitemShut {NoStop}%
\bibitem [{\citenamefont {Lee}\ \emph {et~al.}(2011)\citenamefont {Lee},
  \citenamefont {H\"affner},\ and\ \citenamefont {Cross}}]{PhysRevA.84.031402}%
  \BibitemOpen
  \bibfield  {author} {\bibinfo {author} {\bibfnamefont {T.~E.}\ \bibnamefont
  {Lee}}, \bibinfo {author} {\bibfnamefont {H.}~\bibnamefont {H\"affner}},\
  and\ \bibinfo {author} {\bibfnamefont {M.~C.}\ \bibnamefont {Cross}},\
  }\bibfield  {title} {\bibinfo {title} {Antiferromagnetic phase transition in
  a nonequilibrium lattice of {R}ydberg atoms},\ }\href
  {https://doi.org/10.1103/PhysRevA.84.031402} {\bibfield  {journal} {\bibinfo
  {journal} {Phys. Rev. A}\ }\textbf {\bibinfo {volume} {84}},\ \bibinfo
  {pages} {031402} (\bibinfo {year} {2011})}\BibitemShut {NoStop}%
\bibitem [{\citenamefont {Marcuzzi}\ \emph {et~al.}(2014)\citenamefont
  {Marcuzzi}, \citenamefont {Levi}, \citenamefont {Diehl}, \citenamefont
  {Garrahan},\ and\ \citenamefont {Lesanovsky}}]{PhysRevLett.113.210401}%
  \BibitemOpen
  \bibfield  {author} {\bibinfo {author} {\bibfnamefont {M.}~\bibnamefont
  {Marcuzzi}}, \bibinfo {author} {\bibfnamefont {E.}~\bibnamefont {Levi}},
  \bibinfo {author} {\bibfnamefont {S.}~\bibnamefont {Diehl}}, \bibinfo
  {author} {\bibfnamefont {J.~P.}\ \bibnamefont {Garrahan}},\ and\ \bibinfo
  {author} {\bibfnamefont {I.}~\bibnamefont {Lesanovsky}},\ }\bibfield  {title}
  {\bibinfo {title} {Universal nonequilibrium properties of dissipative
  {R}ydberg gases},\ }\href {https://doi.org/10.1103/PhysRevLett.113.210401}
  {\bibfield  {journal} {\bibinfo  {journal} {Phys. Rev. Lett.}\ }\textbf
  {\bibinfo {volume} {113}},\ \bibinfo {pages} {210401} (\bibinfo {year}
  {2014})}\BibitemShut {NoStop}%
\bibitem [{\citenamefont {Parmee}\ and\ \citenamefont
  {Cooper}(2018)}]{PhysRevA.97.053616}%
  \BibitemOpen
  \bibfield  {author} {\bibinfo {author} {\bibfnamefont {C.~D.}\ \bibnamefont
  {Parmee}}\ and\ \bibinfo {author} {\bibfnamefont {N.~R.}\ \bibnamefont
  {Cooper}},\ }\bibfield  {title} {\bibinfo {title} {Phases of driven two-level
  systems with nonlocal dissipation},\ }\href
  {https://doi.org/10.1103/PhysRevA.97.053616} {\bibfield  {journal} {\bibinfo
  {journal} {Phys. Rev. A}\ }\textbf {\bibinfo {volume} {97}},\ \bibinfo
  {pages} {053616} (\bibinfo {year} {2018})}\BibitemShut {NoStop}%
\bibitem [{\citenamefont {Labouvie}\ \emph {et~al.}(2016)\citenamefont
  {Labouvie}, \citenamefont {Santra}, \citenamefont {Heun},\ and\ \citenamefont
  {Ott}}]{PhysRevLett.116.235302}%
  \BibitemOpen
  \bibfield  {author} {\bibinfo {author} {\bibfnamefont {R.}~\bibnamefont
  {Labouvie}}, \bibinfo {author} {\bibfnamefont {B.}~\bibnamefont {Santra}},
  \bibinfo {author} {\bibfnamefont {S.}~\bibnamefont {Heun}},\ and\ \bibinfo
  {author} {\bibfnamefont {H.}~\bibnamefont {Ott}},\ }\bibfield  {title}
  {\bibinfo {title} {Bistability in a driven-dissipative superfluid},\ }\href
  {https://doi.org/10.1103/PhysRevLett.116.235302} {\bibfield  {journal}
  {\bibinfo  {journal} {Phys. Rev. Lett.}\ }\textbf {\bibinfo {volume} {116}},\
  \bibinfo {pages} {235302} (\bibinfo {year} {2016})}\BibitemShut {NoStop}%
\bibitem [{\citenamefont {Foss-Feig}\ \emph {et~al.}(2017)\citenamefont
  {Foss-Feig}, \citenamefont {Niroula}, \citenamefont {Young}, \citenamefont
  {Hafezi}, \citenamefont {Gorshkov}, \citenamefont {Wilson},\ and\
  \citenamefont {Maghrebi}}]{PhysRevA.95.043826}%
  \BibitemOpen
  \bibfield  {author} {\bibinfo {author} {\bibfnamefont {M.}~\bibnamefont
  {Foss-Feig}}, \bibinfo {author} {\bibfnamefont {P.}~\bibnamefont {Niroula}},
  \bibinfo {author} {\bibfnamefont {J.~T.}\ \bibnamefont {Young}}, \bibinfo
  {author} {\bibfnamefont {M.}~\bibnamefont {Hafezi}}, \bibinfo {author}
  {\bibfnamefont {A.~V.}\ \bibnamefont {Gorshkov}}, \bibinfo {author}
  {\bibfnamefont {R.~M.}\ \bibnamefont {Wilson}},\ and\ \bibinfo {author}
  {\bibfnamefont {M.~F.}\ \bibnamefont {Maghrebi}},\ }\bibfield  {title}
  {\bibinfo {title} {Emergent equilibrium in many-body optical bistability},\
  }\href {https://doi.org/10.1103/PhysRevA.95.043826} {\bibfield  {journal}
  {\bibinfo  {journal} {Phys. Rev. A}\ }\textbf {\bibinfo {volume} {95}},\
  \bibinfo {pages} {043826} (\bibinfo {year} {2017})}\BibitemShut {NoStop}%
\bibitem [{\citenamefont {Biondi}\ \emph {et~al.}(2017)\citenamefont {Biondi},
  \citenamefont {Blatter}, \citenamefont {T\"ureci},\ and\ \citenamefont
  {Schmidt}}]{PhysRevA.96.043809}%
  \BibitemOpen
  \bibfield  {author} {\bibinfo {author} {\bibfnamefont {M.}~\bibnamefont
  {Biondi}}, \bibinfo {author} {\bibfnamefont {G.}~\bibnamefont {Blatter}},
  \bibinfo {author} {\bibfnamefont {H.~E.}\ \bibnamefont {T\"ureci}},\ and\
  \bibinfo {author} {\bibfnamefont {S.}~\bibnamefont {Schmidt}},\ }\bibfield
  {title} {\bibinfo {title} {Nonequilibrium gas-liquid transition in the
  driven-dissipative photonic lattice},\ }\href
  {https://doi.org/10.1103/PhysRevA.96.043809} {\bibfield  {journal} {\bibinfo
  {journal} {Phys. Rev. A}\ }\textbf {\bibinfo {volume} {96}},\ \bibinfo
  {pages} {043809} (\bibinfo {year} {2017})}\BibitemShut {NoStop}%
\bibitem [{\citenamefont {Bon\ifmmode~\check{c}\else \v{c}\fi{}a}\ and\
  \citenamefont {Trugman}(2021)}]{PhysRevB.103.054304}%
  \BibitemOpen
  \bibfield  {author} {\bibinfo {author} {\bibfnamefont {J.}~\bibnamefont
  {Bon\ifmmode~\check{c}\else \v{c}\fi{}a}}\ and\ \bibinfo {author}
  {\bibfnamefont {S.~A.}\ \bibnamefont {Trugman}},\ }\bibfield  {title}
  {\bibinfo {title} {Dynamic properties of a polaron coupled to dispersive
  optical phonons},\ }\href {https://doi.org/10.1103/PhysRevB.103.054304}
  {\bibfield  {journal} {\bibinfo  {journal} {Phys. Rev. B}\ }\textbf {\bibinfo
  {volume} {103}},\ \bibinfo {pages} {054304} (\bibinfo {year}
  {2021})}\BibitemShut {NoStop}%
\bibitem [{\citenamefont {Kennes}\ \emph {et~al.}(2017)\citenamefont {Kennes},
  \citenamefont {Wilner}, \citenamefont {Reichman},\ and\ \citenamefont
  {Millis}}]{Kennes2017}%
  \BibitemOpen
  \bibfield  {author} {\bibinfo {author} {\bibfnamefont {D.~M.}\ \bibnamefont
  {Kennes}}, \bibinfo {author} {\bibfnamefont {E.~Y.}\ \bibnamefont {Wilner}},
  \bibinfo {author} {\bibfnamefont {D.~R.}\ \bibnamefont {Reichman}},\ and\
  \bibinfo {author} {\bibfnamefont {A.~J.}\ \bibnamefont {Millis}},\ }\bibfield
   {title} {\bibinfo {title} {Transient superconductivity from electronic
  squeezing of optically pumped phonons},\ }\href
  {https://doi.org/10.1038/nphys4024} {\bibfield  {journal} {\bibinfo
  {journal} {Nature Physics}\ }\textbf {\bibinfo {volume} {13}},\ \bibinfo
  {pages} {479} (\bibinfo {year} {2017})}\BibitemShut {NoStop}%
\bibitem [{\citenamefont {Sous}\ \emph {et~al.}(2021)\citenamefont {Sous},
  \citenamefont {Kloss}, \citenamefont {Kennes}, \citenamefont {Reichman},\
  and\ \citenamefont {Millis}}]{Sous2021}%
  \BibitemOpen
  \bibfield  {author} {\bibinfo {author} {\bibfnamefont {J.}~\bibnamefont
  {Sous}}, \bibinfo {author} {\bibfnamefont {B.}~\bibnamefont {Kloss}},
  \bibinfo {author} {\bibfnamefont {D.~M.}\ \bibnamefont {Kennes}}, \bibinfo
  {author} {\bibfnamefont {D.~R.}\ \bibnamefont {Reichman}},\ and\ \bibinfo
  {author} {\bibfnamefont {A.~J.}\ \bibnamefont {Millis}},\ }\bibfield  {title}
  {\bibinfo {title} {Phonon-induced disorder in dynamics of optically pumped
  metals from nonlinear electron-phonon coupling},\ }\href
  {https://doi.org/10.1038/s41467-021-26030-3} {\bibfield  {journal} {\bibinfo
  {journal} {Nature Communications}\ }\textbf {\bibinfo {volume} {12}},\
  \bibinfo {pages} {5803} (\bibinfo {year} {2021})}\BibitemShut {NoStop}%
\bibitem [{SM()}]{SM}%
  \BibitemOpen
  \href@noop {} {\bibinfo  {journal} {See the Supplemental Materials at
  http://xxx.yyy.zzz for further details of the model Hamiltonian,
  corresponding equations of motion, resonance/off-resonance phononic and
  electronic responses, and energy flows}\ }\BibitemShut {NoStop}%
\bibitem [{\citenamefont {Kim}\ \emph {et~al.}(2003)\citenamefont {Kim},
  \citenamefont {Kordyuk}, \citenamefont {Borisenko}, \citenamefont {Koitzsch},
  \citenamefont {Knupfer}, \citenamefont {Berger},\ and\ \citenamefont
  {Fink}}]{PhysRevLett.91.167002}%
  \BibitemOpen
\bibfield  {journal} {  }\bibfield  {author} {\bibinfo {author} {\bibfnamefont
  {T.~K.}\ \bibnamefont {Kim}}, \bibinfo {author} {\bibfnamefont {A.~A.}\
  \bibnamefont {Kordyuk}}, \bibinfo {author} {\bibfnamefont {S.~V.}\
  \bibnamefont {Borisenko}}, \bibinfo {author} {\bibfnamefont {A.}~\bibnamefont
  {Koitzsch}}, \bibinfo {author} {\bibfnamefont {M.}~\bibnamefont {Knupfer}},
  \bibinfo {author} {\bibfnamefont {H.}~\bibnamefont {Berger}},\ and\ \bibinfo
  {author} {\bibfnamefont {J.}~\bibnamefont {Fink}},\ }\bibfield  {title}
  {\bibinfo {title} {Doping dependence of the mass enhancement in
  ({Pb},{Bi})$_2${Sr}$_2${CaCu}$_2${O}$_8$ at the antinodal point in the
  superconducting and normal states},\ }\href
  {https://doi.org/10.1103/PhysRevLett.91.167002} {\bibfield  {journal}
  {\bibinfo  {journal} {Phys. Rev. Lett.}\ }\textbf {\bibinfo {volume} {91}},\
  \bibinfo {pages} {167002} (\bibinfo {year} {2003})}\BibitemShut {NoStop}%
\bibitem [{\citenamefont {Sentef}(2017)}]{PhysRevB.95.205111}%
  \BibitemOpen
  \bibfield  {author} {\bibinfo {author} {\bibfnamefont {M.~A.}\ \bibnamefont
  {Sentef}},\ }\bibfield  {title} {\bibinfo {title} {Light-enhanced
  electron-phonon coupling from nonlinear electron-phonon coupling},\ }\href
  {https://doi.org/10.1103/PhysRevB.95.205111} {\bibfield  {journal} {\bibinfo
  {journal} {Phys. Rev. B}\ }\textbf {\bibinfo {volume} {95}},\ \bibinfo
  {pages} {205111} (\bibinfo {year} {2017})}\BibitemShut {NoStop}%
\bibitem [{\citenamefont {Puviani}\ and\ \citenamefont
  {Sentef}(2018)}]{PhysRevB.98.165138}%
  \BibitemOpen
  \bibfield  {author} {\bibinfo {author} {\bibfnamefont {M.}~\bibnamefont
  {Puviani}}\ and\ \bibinfo {author} {\bibfnamefont {M.~A.}\ \bibnamefont
  {Sentef}},\ }\bibfield  {title} {\bibinfo {title} {Quantum nonlinear
  phononics route towards nonequilibrium materials engineering: Melting
  dynamics of a ferrielectric charge density wave},\ }\href
  {https://doi.org/10.1103/PhysRevB.98.165138} {\bibfield  {journal} {\bibinfo
  {journal} {Phys. Rev. B}\ }\textbf {\bibinfo {volume} {98}},\ \bibinfo
  {pages} {165138} (\bibinfo {year} {2018})}\BibitemShut {NoStop}%
\bibitem [{\citenamefont {Lindblad}(1976)}]{lindblad1976}%
  \BibitemOpen
  \bibfield  {author} {\bibinfo {author} {\bibfnamefont {G.}~\bibnamefont
  {Lindblad}},\ }\bibfield  {title} {\bibinfo {title} {On the generators of
  quantum dynamical semigroups},\ }\href
  {https://projecteuclid.org:443/euclid.cmp/1103899849} {\bibfield  {journal}
  {\bibinfo  {journal} {Comm. Math. Phys.}\ }\textbf {\bibinfo {volume} {48}},\
  \bibinfo {pages} {119} (\bibinfo {year} {1976})}\BibitemShut {NoStop}%
\bibitem [{\citenamefont {Breuer}\ and\ \citenamefont
  {Petruccione}(2007)}]{breuer2007theory}%
  \BibitemOpen
  \bibfield  {author} {\bibinfo {author} {\bibfnamefont {H.}~\bibnamefont
  {Breuer}}\ and\ \bibinfo {author} {\bibfnamefont {F.}~\bibnamefont
  {Petruccione}},\ }\href {https://books.google.de/books?id=DkcJPwAACAAJ}
  {\emph {\bibinfo {title} {The Theory of Open Quantum Systems}}}\ (\bibinfo
  {publisher} {OUP Oxford},\ \bibinfo {year} {2007})\BibitemShut {NoStop}%
\bibitem [{\citenamefont {Mitrano}\ \emph {et~al.}(2014)\citenamefont
  {Mitrano}, \citenamefont {Cotugno}, \citenamefont {Clark}, \citenamefont
  {Singla}, \citenamefont {Kaiser}, \citenamefont {St\"ahler}, \citenamefont
  {Beyer}, \citenamefont {Dressel}, \citenamefont {Baldassarre}, \citenamefont
  {Nicoletti}, \citenamefont {Perucchi}, \citenamefont {Hasegawa},
  \citenamefont {Okamoto}, \citenamefont {Jaksch},\ and\ \citenamefont
  {Cavalleri}}]{PhysRevLett.112.117801}%
  \BibitemOpen
  \bibfield  {author} {\bibinfo {author} {\bibfnamefont {M.}~\bibnamefont
  {Mitrano}}, \bibinfo {author} {\bibfnamefont {G.}~\bibnamefont {Cotugno}},
  \bibinfo {author} {\bibfnamefont {S.~R.}\ \bibnamefont {Clark}}, \bibinfo
  {author} {\bibfnamefont {R.}~\bibnamefont {Singla}}, \bibinfo {author}
  {\bibfnamefont {S.}~\bibnamefont {Kaiser}}, \bibinfo {author} {\bibfnamefont
  {J.}~\bibnamefont {St\"ahler}}, \bibinfo {author} {\bibfnamefont
  {R.}~\bibnamefont {Beyer}}, \bibinfo {author} {\bibfnamefont
  {M.}~\bibnamefont {Dressel}}, \bibinfo {author} {\bibfnamefont
  {L.}~\bibnamefont {Baldassarre}}, \bibinfo {author} {\bibfnamefont
  {D.}~\bibnamefont {Nicoletti}}, \bibinfo {author} {\bibfnamefont
  {A.}~\bibnamefont {Perucchi}}, \bibinfo {author} {\bibfnamefont
  {T.}~\bibnamefont {Hasegawa}}, \bibinfo {author} {\bibfnamefont
  {H.}~\bibnamefont {Okamoto}}, \bibinfo {author} {\bibfnamefont
  {D.}~\bibnamefont {Jaksch}},\ and\ \bibinfo {author} {\bibfnamefont
  {A.}~\bibnamefont {Cavalleri}},\ }\bibfield  {title} {\bibinfo {title}
  {Pressure-dependent relaxation in the photoexcited mott insulator
  {ET}$-${F}$_2${TCNQ}: Influence of hopping and correlations on quasiparticle
  recombination rates},\ }\href
  {https://doi.org/10.1103/PhysRevLett.112.117801} {\bibfield  {journal}
  {\bibinfo  {journal} {Phys. Rev. Lett.}\ }\textbf {\bibinfo {volume} {112}},\
  \bibinfo {pages} {117801} (\bibinfo {year} {2014})}\BibitemShut {NoStop}%
\bibitem [{Note1()}]{Note1}%
  \BibitemOpen
  \bibinfo {note} {Note that the phonon relaxation is not modified because it
  remains gapped throughout the simulations at zero temperature.}\BibitemShut
  {Stop}%
\bibitem [{\citenamefont {Duan-Ming}\ \emph {et~al.}(2007)\citenamefont
  {Duan-Ming}, \citenamefont {Ran-Ran}, \citenamefont {Zhi-Hua}, \citenamefont
  {Li}, \citenamefont {Li}, \citenamefont {Xin-Yu}, \citenamefont {Dan},
  \citenamefont {Gao-Bin},\ and\ \citenamefont {De-Zhi}}]{Zhang7}%
  \BibitemOpen
  \bibfield  {author} {\bibinfo {author} {\bibfnamefont {Z.}~\bibnamefont
  {Duan-Ming}}, \bibinfo {author} {\bibfnamefont {F.}~\bibnamefont {Ran-Ran}},
  \bibinfo {author} {\bibfnamefont {L.}~\bibnamefont {Zhi-Hua}}, \bibinfo
  {author} {\bibfnamefont {G.}~\bibnamefont {Li}}, \bibinfo {author}
  {\bibfnamefont {L.}~\bibnamefont {Li}}, \bibinfo {author} {\bibfnamefont
  {T.}~\bibnamefont {Xin-Yu}}, \bibinfo {author} {\bibfnamefont
  {L.}~\bibnamefont {Dan}}, \bibinfo {author} {\bibfnamefont {L.}~\bibnamefont
  {Gao-Bin}},\ and\ \bibinfo {author} {\bibfnamefont {H.}~\bibnamefont
  {De-Zhi}},\ }\bibfield  {title} {\bibinfo {title} {A new synthetical model of
  high-power pulsed laser ablation},\ }\href
  {https://doi.org/10.1088/0253-6102/48/1/033} {\bibfield  {journal} {\bibinfo
  {journal} {Communications in Theoretical Physics}\ }\textbf {\bibinfo
  {volume} {48}},\ \bibinfo {pages} {163} (\bibinfo {year} {2007})}\BibitemShut
  {NoStop}%
\bibitem [{\citenamefont {Rowe}\ \emph {et~al.}(2023)\citenamefont {Rowe},
  \citenamefont {Yuan}, \citenamefont {Buzzi}, \citenamefont {Jotzu},
  \citenamefont {Zhu}, \citenamefont {Fechner}, \citenamefont {Först},
  \citenamefont {Liu}, \citenamefont {Pontiroli}, \citenamefont {Riccò},\ and\
  \citenamefont {Cavalleri}}]{https://doi.org/10.48550/arxiv.2301.08633}%
  \BibitemOpen
  \bibfield  {author} {\bibinfo {author} {\bibfnamefont {E.}~\bibnamefont
  {Rowe}}, \bibinfo {author} {\bibfnamefont {B.}~\bibnamefont {Yuan}}, \bibinfo
  {author} {\bibfnamefont {M.}~\bibnamefont {Buzzi}}, \bibinfo {author}
  {\bibfnamefont {G.}~\bibnamefont {Jotzu}}, \bibinfo {author} {\bibfnamefont
  {Y.}~\bibnamefont {Zhu}}, \bibinfo {author} {\bibfnamefont {M.}~\bibnamefont
  {Fechner}}, \bibinfo {author} {\bibfnamefont {M.}~\bibnamefont {Först}},
  \bibinfo {author} {\bibfnamefont {B.}~\bibnamefont {Liu}}, \bibinfo {author}
  {\bibfnamefont {D.}~\bibnamefont {Pontiroli}}, \bibinfo {author}
  {\bibfnamefont {M.}~\bibnamefont {Riccò}},\ and\ \bibinfo {author}
  {\bibfnamefont {A.}~\bibnamefont {Cavalleri}},\ }\href
  {https://doi.org/10.48550/ARXIV.2301.08633} {\bibinfo {title} {Giant resonant
  enhancement for photo-induced superconductivity in {K}$_{3}${C}$_{60}$}}
  (\bibinfo {year} {2023})\BibitemShut {NoStop}%
\bibitem [{\citenamefont {Wang}\ \emph {et~al.}(2023)\citenamefont {Wang},
  \citenamefont {Adelinia}, \citenamefont {Chavez-Cervantes}, \citenamefont
  {Matsuyama}, \citenamefont {Fechner}, \citenamefont {Buzzi}, \citenamefont
  {Meier},\ and\ \citenamefont
  {Cavalleri}}]{https://doi.org/10.48550/arxiv.2301.06425}%
  \BibitemOpen
  \bibfield  {author} {\bibinfo {author} {\bibfnamefont {E.}~\bibnamefont
  {Wang}}, \bibinfo {author} {\bibfnamefont {J.~D.}\ \bibnamefont {Adelinia}},
  \bibinfo {author} {\bibfnamefont {M.}~\bibnamefont {Chavez-Cervantes}},
  \bibinfo {author} {\bibfnamefont {T.}~\bibnamefont {Matsuyama}}, \bibinfo
  {author} {\bibfnamefont {M.}~\bibnamefont {Fechner}}, \bibinfo {author}
  {\bibfnamefont {M.}~\bibnamefont {Buzzi}}, \bibinfo {author} {\bibfnamefont
  {G.}~\bibnamefont {Meier}},\ and\ \bibinfo {author} {\bibfnamefont
  {A.}~\bibnamefont {Cavalleri}},\ }\href
  {https://doi.org/10.48550/ARXIV.2301.06425} {\bibinfo {title} {Nonlinear
  transport in a photo-induced superconductor}} (\bibinfo {year}
  {2023})\BibitemShut {NoStop}%
\bibitem [{\citenamefont {Dolgirev}\ \emph {et~al.}(2022)\citenamefont
  {Dolgirev}, \citenamefont {Zong}, \citenamefont {Michael}, \citenamefont
  {Curtis}, \citenamefont {Podolsky}, \citenamefont {Cavalleri},\ and\
  \citenamefont {Demler}}]{Dolgirev2022}%
  \BibitemOpen
  \bibfield  {author} {\bibinfo {author} {\bibfnamefont {P.~E.}\ \bibnamefont
  {Dolgirev}}, \bibinfo {author} {\bibfnamefont {A.}~\bibnamefont {Zong}},
  \bibinfo {author} {\bibfnamefont {M.~H.}\ \bibnamefont {Michael}}, \bibinfo
  {author} {\bibfnamefont {J.~B.}\ \bibnamefont {Curtis}}, \bibinfo {author}
  {\bibfnamefont {D.}~\bibnamefont {Podolsky}}, \bibinfo {author}
  {\bibfnamefont {A.}~\bibnamefont {Cavalleri}},\ and\ \bibinfo {author}
  {\bibfnamefont {E.}~\bibnamefont {Demler}},\ }\bibfield  {title} {\bibinfo
  {title} {Periodic dynamics in superconductors induced by an impulsive optical
  quench},\ }\href {https://doi.org/10.1038/s42005-022-01007-w} {\bibfield
  {journal} {\bibinfo  {journal} {Communications Physics}\ }\textbf {\bibinfo
  {volume} {5}},\ \bibinfo {pages} {234} (\bibinfo {year} {2022})}\BibitemShut
  {NoStop}%
\bibitem [{\citenamefont {von Hoegen}\ \emph {et~al.}(2022)\citenamefont {von
  Hoegen}, \citenamefont {Fechner}, \citenamefont {F\"orst}, \citenamefont
  {Taherian}, \citenamefont {Rowe}, \citenamefont {Ribak}, \citenamefont
  {Porras}, \citenamefont {Keimer}, \citenamefont {Michael}, \citenamefont
  {Demler},\ and\ \citenamefont {Cavalleri}}]{PhysRevX.12.031008}%
  \BibitemOpen
  \bibfield  {author} {\bibinfo {author} {\bibfnamefont {A.}~\bibnamefont {von
  Hoegen}}, \bibinfo {author} {\bibfnamefont {M.}~\bibnamefont {Fechner}},
  \bibinfo {author} {\bibfnamefont {M.}~\bibnamefont {F\"orst}}, \bibinfo
  {author} {\bibfnamefont {N.}~\bibnamefont {Taherian}}, \bibinfo {author}
  {\bibfnamefont {E.}~\bibnamefont {Rowe}}, \bibinfo {author} {\bibfnamefont
  {A.}~\bibnamefont {Ribak}}, \bibinfo {author} {\bibfnamefont
  {J.}~\bibnamefont {Porras}}, \bibinfo {author} {\bibfnamefont
  {B.}~\bibnamefont {Keimer}}, \bibinfo {author} {\bibfnamefont
  {M.}~\bibnamefont {Michael}}, \bibinfo {author} {\bibfnamefont
  {E.}~\bibnamefont {Demler}},\ and\ \bibinfo {author} {\bibfnamefont
  {A.}~\bibnamefont {Cavalleri}},\ }\bibfield  {title} {\bibinfo {title}
  {Amplification of superconducting fluctuations in driven
  {YBa}$_2${Cu}$_3${O}$_{6+x}$},\ }\href
  {https://doi.org/10.1103/PhysRevX.12.031008} {\bibfield  {journal} {\bibinfo
  {journal} {Phys. Rev. X}\ }\textbf {\bibinfo {volume} {12}},\ \bibinfo
  {pages} {031008} (\bibinfo {year} {2022})}\BibitemShut {NoStop}%
\bibitem [{\citenamefont {Henstridge}\ \emph {et~al.}(2022)\citenamefont
  {Henstridge}, \citenamefont {F{\"o}rst}, \citenamefont {Rowe}, \citenamefont
  {Fechner},\ and\ \citenamefont {Cavalleri}}]{Henstridge2022}%
  \BibitemOpen
  \bibfield  {author} {\bibinfo {author} {\bibfnamefont {M.}~\bibnamefont
  {Henstridge}}, \bibinfo {author} {\bibfnamefont {M.}~\bibnamefont
  {F{\"o}rst}}, \bibinfo {author} {\bibfnamefont {E.}~\bibnamefont {Rowe}},
  \bibinfo {author} {\bibfnamefont {M.}~\bibnamefont {Fechner}},\ and\ \bibinfo
  {author} {\bibfnamefont {A.}~\bibnamefont {Cavalleri}},\ }\bibfield  {title}
  {\bibinfo {title} {Nonlocal nonlinear phononics},\ }\href
  {https://doi.org/10.1038/s41567-022-01512-3} {\bibfield  {journal} {\bibinfo
  {journal} {Nature Physics}\ }\textbf {\bibinfo {volume} {18}},\ \bibinfo
  {pages} {457} (\bibinfo {year} {2022})}\BibitemShut {NoStop}%
\bibitem [{\citenamefont {Buzzi}\ \emph
  {et~al.}(2021{\natexlab{a}})\citenamefont {Buzzi}, \citenamefont {Nicoletti},
  \citenamefont {Fava}, \citenamefont {Jotzu}, \citenamefont {Miyagawa},
  \citenamefont {Kanoda}, \citenamefont {Henderson}, \citenamefont {Siegrist},
  \citenamefont {Schlueter}, \citenamefont {Nam}, \citenamefont {Ardavan},\
  and\ \citenamefont {Cavalleri}}]{PhysRevLett.127.197002}%
  \BibitemOpen
  \bibfield  {author} {\bibinfo {author} {\bibfnamefont {M.}~\bibnamefont
  {Buzzi}}, \bibinfo {author} {\bibfnamefont {D.}~\bibnamefont {Nicoletti}},
  \bibinfo {author} {\bibfnamefont {S.}~\bibnamefont {Fava}}, \bibinfo {author}
  {\bibfnamefont {G.}~\bibnamefont {Jotzu}}, \bibinfo {author} {\bibfnamefont
  {K.}~\bibnamefont {Miyagawa}}, \bibinfo {author} {\bibfnamefont
  {K.}~\bibnamefont {Kanoda}}, \bibinfo {author} {\bibfnamefont
  {A.}~\bibnamefont {Henderson}}, \bibinfo {author} {\bibfnamefont
  {T.}~\bibnamefont {Siegrist}}, \bibinfo {author} {\bibfnamefont {J.~A.}\
  \bibnamefont {Schlueter}}, \bibinfo {author} {\bibfnamefont {M.-S.}\
  \bibnamefont {Nam}}, \bibinfo {author} {\bibfnamefont {A.}~\bibnamefont
  {Ardavan}},\ and\ \bibinfo {author} {\bibfnamefont {A.}~\bibnamefont
  {Cavalleri}},\ }\bibfield  {title} {\bibinfo {title} {Phase diagram for
  light-induced superconductivity in $\kappa-$({ET})$_{2}-${X}},\ }\href
  {https://doi.org/10.1103/PhysRevLett.127.197002} {\bibfield  {journal}
  {\bibinfo  {journal} {Phys. Rev. Lett.}\ }\textbf {\bibinfo {volume} {127}},\
  \bibinfo {pages} {197002} (\bibinfo {year} {2021}{\natexlab{a}})}\BibitemShut
  {NoStop}%
\bibitem [{\citenamefont {Buzzi}\ \emph
  {et~al.}(2021{\natexlab{b}})\citenamefont {Buzzi}, \citenamefont {Jotzu},
  \citenamefont {Cavalleri}, \citenamefont {Cirac}, \citenamefont {Demler},
  \citenamefont {Halperin}, \citenamefont {Lukin}, \citenamefont {Shi},
  \citenamefont {Wang},\ and\ \citenamefont {Podolsky}}]{PhysRevX.11.011055}%
  \BibitemOpen
  \bibfield  {author} {\bibinfo {author} {\bibfnamefont {M.}~\bibnamefont
  {Buzzi}}, \bibinfo {author} {\bibfnamefont {G.}~\bibnamefont {Jotzu}},
  \bibinfo {author} {\bibfnamefont {A.}~\bibnamefont {Cavalleri}}, \bibinfo
  {author} {\bibfnamefont {J.~I.}\ \bibnamefont {Cirac}}, \bibinfo {author}
  {\bibfnamefont {E.~A.}\ \bibnamefont {Demler}}, \bibinfo {author}
  {\bibfnamefont {B.~I.}\ \bibnamefont {Halperin}}, \bibinfo {author}
  {\bibfnamefont {M.~D.}\ \bibnamefont {Lukin}}, \bibinfo {author}
  {\bibfnamefont {T.}~\bibnamefont {Shi}}, \bibinfo {author} {\bibfnamefont
  {Y.}~\bibnamefont {Wang}},\ and\ \bibinfo {author} {\bibfnamefont
  {D.}~\bibnamefont {Podolsky}},\ }\bibfield  {title} {\bibinfo {title}
  {Higgs-mediated optical amplification in a nonequilibrium superconductor},\
  }\href {https://doi.org/10.1103/PhysRevX.11.011055} {\bibfield  {journal}
  {\bibinfo  {journal} {Phys. Rev. X}\ }\textbf {\bibinfo {volume} {11}},\
  \bibinfo {pages} {011055} (\bibinfo {year} {2021}{\natexlab{b}})}\BibitemShut
  {NoStop}%
\bibitem [{\citenamefont {Buzzi}\ \emph {et~al.}(2020)\citenamefont {Buzzi},
  \citenamefont {Nicoletti}, \citenamefont {Fechner}, \citenamefont
  {Tancogne-Dejean}, \citenamefont {Sentef}, \citenamefont {Georges},
  \citenamefont {Biesner}, \citenamefont {Uykur}, \citenamefont {Dressel},
  \citenamefont {Henderson}, \citenamefont {Siegrist}, \citenamefont
  {Schlueter}, \citenamefont {Miyagawa}, \citenamefont {Kanoda}, \citenamefont
  {Nam}, \citenamefont {Ardavan}, \citenamefont {Coulthard}, \citenamefont
  {Tindall}, \citenamefont {Schlawin}, \citenamefont {Jaksch},\ and\
  \citenamefont {Cavalleri}}]{PhysRevX.10.031028}%
  \BibitemOpen
  \bibfield  {author} {\bibinfo {author} {\bibfnamefont {M.}~\bibnamefont
  {Buzzi}}, \bibinfo {author} {\bibfnamefont {D.}~\bibnamefont {Nicoletti}},
  \bibinfo {author} {\bibfnamefont {M.}~\bibnamefont {Fechner}}, \bibinfo
  {author} {\bibfnamefont {N.}~\bibnamefont {Tancogne-Dejean}}, \bibinfo
  {author} {\bibfnamefont {M.~A.}\ \bibnamefont {Sentef}}, \bibinfo {author}
  {\bibfnamefont {A.}~\bibnamefont {Georges}}, \bibinfo {author} {\bibfnamefont
  {T.}~\bibnamefont {Biesner}}, \bibinfo {author} {\bibfnamefont
  {E.}~\bibnamefont {Uykur}}, \bibinfo {author} {\bibfnamefont
  {M.}~\bibnamefont {Dressel}}, \bibinfo {author} {\bibfnamefont
  {A.}~\bibnamefont {Henderson}}, \bibinfo {author} {\bibfnamefont
  {T.}~\bibnamefont {Siegrist}}, \bibinfo {author} {\bibfnamefont {J.~A.}\
  \bibnamefont {Schlueter}}, \bibinfo {author} {\bibfnamefont {K.}~\bibnamefont
  {Miyagawa}}, \bibinfo {author} {\bibfnamefont {K.}~\bibnamefont {Kanoda}},
  \bibinfo {author} {\bibfnamefont {M.-S.}\ \bibnamefont {Nam}}, \bibinfo
  {author} {\bibfnamefont {A.}~\bibnamefont {Ardavan}}, \bibinfo {author}
  {\bibfnamefont {J.}~\bibnamefont {Coulthard}}, \bibinfo {author}
  {\bibfnamefont {J.}~\bibnamefont {Tindall}}, \bibinfo {author} {\bibfnamefont
  {F.}~\bibnamefont {Schlawin}}, \bibinfo {author} {\bibfnamefont
  {D.}~\bibnamefont {Jaksch}},\ and\ \bibinfo {author} {\bibfnamefont
  {A.}~\bibnamefont {Cavalleri}},\ }\bibfield  {title} {\bibinfo {title}
  {Photomolecular high-temperature superconductivity},\ }\href
  {https://doi.org/10.1103/PhysRevX.10.031028} {\bibfield  {journal} {\bibinfo
  {journal} {Phys. Rev. X}\ }\textbf {\bibinfo {volume} {10}},\ \bibinfo
  {pages} {031028} (\bibinfo {year} {2020})}\BibitemShut {NoStop}%
\bibitem [{\citenamefont {Dann}\ \emph {et~al.}(2018)\citenamefont {Dann},
  \citenamefont {Levy},\ and\ \citenamefont {Kosloff}}]{PhysRevA.98.052129}%
  \BibitemOpen
  \bibfield  {author} {\bibinfo {author} {\bibfnamefont {R.}~\bibnamefont
  {Dann}}, \bibinfo {author} {\bibfnamefont {A.}~\bibnamefont {Levy}},\ and\
  \bibinfo {author} {\bibfnamefont {R.}~\bibnamefont {Kosloff}},\ }\bibfield
  {title} {\bibinfo {title} {Time-dependent {M}arkovian quantum master
  equation},\ }\href {https://doi.org/10.1103/PhysRevA.98.052129} {\bibfield
  {journal} {\bibinfo  {journal} {Phys. Rev. A}\ }\textbf {\bibinfo {volume}
  {98}},\ \bibinfo {pages} {052129} (\bibinfo {year} {2018})}\BibitemShut
  {NoStop}%
\bibitem [{\citenamefont {Johnson}\ \emph {et~al.}(2009)\citenamefont
  {Johnson}, \citenamefont {Beaud}, \citenamefont {Vorobeva}, \citenamefont
  {Milne}, \citenamefont {Murray}, \citenamefont {Fahy},\ and\ \citenamefont
  {Ingold}}]{PhysRevLett.102.175503}%
  \BibitemOpen
  \bibfield  {author} {\bibinfo {author} {\bibfnamefont {S.~L.}\ \bibnamefont
  {Johnson}}, \bibinfo {author} {\bibfnamefont {P.}~\bibnamefont {Beaud}},
  \bibinfo {author} {\bibfnamefont {E.}~\bibnamefont {Vorobeva}}, \bibinfo
  {author} {\bibfnamefont {C.~J.}\ \bibnamefont {Milne}}, \bibinfo {author}
  {\bibfnamefont {E.~D.}\ \bibnamefont {Murray}}, \bibinfo {author}
  {\bibfnamefont {S.}~\bibnamefont {Fahy}},\ and\ \bibinfo {author}
  {\bibfnamefont {G.}~\bibnamefont {Ingold}},\ }\bibfield  {title} {\bibinfo
  {title} {Directly observing squeezed phonon states with femtosecond x-ray
  diffraction},\ }\href {https://doi.org/10.1103/PhysRevLett.102.175503}
  {\bibfield  {journal} {\bibinfo  {journal} {Phys. Rev. Lett.}\ }\textbf
  {\bibinfo {volume} {102}},\ \bibinfo {pages} {175503} (\bibinfo {year}
  {2009})}\BibitemShut {NoStop}%
\bibitem [{\citenamefont {Lindemann}(1910)}]{Lindemann1910}%
  \BibitemOpen
  \bibfield  {author} {\bibinfo {author} {\bibfnamefont {F.~A.}\ \bibnamefont
  {Lindemann}},\ }\bibfield  {title} {\bibinfo {title} {The calculation of
  molecular vibration frequencies},\ }\href
  {https://ntrs.nasa.gov/citations/19840027015} {\bibfield  {journal} {\bibinfo
   {journal} {Phys. Z.}\ }\textbf {\bibinfo {volume} {11}},\ \bibinfo {pages}
  {609} (\bibinfo {year} {1910})}\BibitemShut {NoStop}%
\bibitem [{\citenamefont {Seipt}\ \emph {et~al.}(2019)\citenamefont {Seipt},
  \citenamefont {Kharin},\ and\ \citenamefont
  {Rykovanov}}]{PhysRevLett.122.204802}%
  \BibitemOpen
  \bibfield  {author} {\bibinfo {author} {\bibfnamefont {D.}~\bibnamefont
  {Seipt}}, \bibinfo {author} {\bibfnamefont {V.~Y.}\ \bibnamefont {Kharin}},\
  and\ \bibinfo {author} {\bibfnamefont {S.~G.}\ \bibnamefont {Rykovanov}},\
  }\bibfield  {title} {\bibinfo {title} {Optimizing laser pulses for
  narrow-band inverse {C}ompton sources in the high-intensity regime},\ }\href
  {https://doi.org/10.1103/PhysRevLett.122.204802} {\bibfield  {journal}
  {\bibinfo  {journal} {Phys. Rev. Lett.}\ }\textbf {\bibinfo {volume} {122}},\
  \bibinfo {pages} {204802} (\bibinfo {year} {2019})}\BibitemShut {NoStop}%
\bibitem [{\citenamefont {Kharin}\ \emph {et~al.}(2018)\citenamefont {Kharin},
  \citenamefont {Seipt},\ and\ \citenamefont
  {Rykovanov}}]{PhysRevLett.120.044802}%
  \BibitemOpen
  \bibfield  {author} {\bibinfo {author} {\bibfnamefont {V.~Y.}\ \bibnamefont
  {Kharin}}, \bibinfo {author} {\bibfnamefont {D.}~\bibnamefont {Seipt}},\ and\
  \bibinfo {author} {\bibfnamefont {S.~G.}\ \bibnamefont {Rykovanov}},\
  }\bibfield  {title} {\bibinfo {title} {Higher-dimensional caustics in
  nonlinear {C}ompton scattering},\ }\href
  {https://doi.org/10.1103/PhysRevLett.120.044802} {\bibfield  {journal}
  {\bibinfo  {journal} {Phys. Rev. Lett.}\ }\textbf {\bibinfo {volume} {120}},\
  \bibinfo {pages} {044802} (\bibinfo {year} {2018})}\BibitemShut {NoStop}%
\bibitem [{\citenamefont {Eckhardt}\ \emph {et~al.}(2023)\citenamefont
  {Eckhardt}, \citenamefont {Chattopadhyay}, \citenamefont {Kennes},
  \citenamefont {Demler}, \citenamefont {Sentef},\ and\ \citenamefont
  {Michael}}]{eckhardt2023theory}%
  \BibitemOpen
  \bibfield  {author} {\bibinfo {author} {\bibfnamefont {C.~J.}\ \bibnamefont
  {Eckhardt}}, \bibinfo {author} {\bibfnamefont {S.}~\bibnamefont
  {Chattopadhyay}}, \bibinfo {author} {\bibfnamefont {D.~M.}\ \bibnamefont
  {Kennes}}, \bibinfo {author} {\bibfnamefont {E.~A.}\ \bibnamefont {Demler}},
  \bibinfo {author} {\bibfnamefont {M.~A.}\ \bibnamefont {Sentef}},\ and\
  \bibinfo {author} {\bibfnamefont {M.~H.}\ \bibnamefont {Michael}},\ }\href
  {https://doi.org/10.48550/arXiv.2303.02176} {\bibinfo {title} {Theory of
  resonantly enhanced photo-induced superconductivity}} (\bibinfo {year}
  {2023})\BibitemShut {NoStop}%
\bibitem [{\citenamefont {Chattopadhyay}\ \emph {et~al.}(2023)\citenamefont
  {Chattopadhyay}, \citenamefont {Eckhardt}, \citenamefont {Kennes},
  \citenamefont {Sentef}, \citenamefont {Shin}, \citenamefont {Rubio},
  \citenamefont {Cavalleri}, \citenamefont {Demler},\ and\ \citenamefont
  {Michael}}]{chattopadhyay2023mechanisms}%
  \BibitemOpen
  \bibfield  {author} {\bibinfo {author} {\bibfnamefont {S.}~\bibnamefont
  {Chattopadhyay}}, \bibinfo {author} {\bibfnamefont {C.~J.}\ \bibnamefont
  {Eckhardt}}, \bibinfo {author} {\bibfnamefont {D.~M.}\ \bibnamefont
  {Kennes}}, \bibinfo {author} {\bibfnamefont {M.~A.}\ \bibnamefont {Sentef}},
  \bibinfo {author} {\bibfnamefont {D.}~\bibnamefont {Shin}}, \bibinfo {author}
  {\bibfnamefont {A.}~\bibnamefont {Rubio}}, \bibinfo {author} {\bibfnamefont
  {A.}~\bibnamefont {Cavalleri}}, \bibinfo {author} {\bibfnamefont {E.~A.}\
  \bibnamefont {Demler}},\ and\ \bibinfo {author} {\bibfnamefont {M.~H.}\
  \bibnamefont {Michael}},\ }\href {https://doi.org/10.48550/arXiv.2303.15355}
  {\bibinfo {title} {Mechanisms for long-lived, photo-induced
  superconductivity}} (\bibinfo {year} {2023})\BibitemShut {NoStop}%
\bibitem [{\citenamefont {Schlawin}\ \emph {et~al.}(2022)\citenamefont
  {Schlawin}, \citenamefont {Kennes},\ and\ \citenamefont
  {Sentef}}]{Schlawin2022}%
  \BibitemOpen
  \bibfield  {author} {\bibinfo {author} {\bibfnamefont {F.}~\bibnamefont
  {Schlawin}}, \bibinfo {author} {\bibfnamefont {D.~M.}\ \bibnamefont
  {Kennes}},\ and\ \bibinfo {author} {\bibfnamefont {M.~A.}\ \bibnamefont
  {Sentef}},\ }\bibfield  {title} {\bibinfo {title} {Cavity quantum
  materials},\ }\href {https://doi.org/10.1063/5.0083825} {\bibfield  {journal}
  {\bibinfo  {journal} {Applied Physics Reviews}\ }\textbf {\bibinfo {volume}
  {9}},\ \bibinfo {pages} {011312} (\bibinfo {year} {2022})}\BibitemShut
  {NoStop}%
\bibitem [{\citenamefont {Sentef}\ \emph {et~al.}(2018)\citenamefont {Sentef},
  \citenamefont {Ruggenthaler},\ and\ \citenamefont
  {Rubio}}]{doi:10.1126/sciadv.aau6969}%
  \BibitemOpen
  \bibfield  {author} {\bibinfo {author} {\bibfnamefont {M.~A.}\ \bibnamefont
  {Sentef}}, \bibinfo {author} {\bibfnamefont {M.}~\bibnamefont
  {Ruggenthaler}},\ and\ \bibinfo {author} {\bibfnamefont {A.}~\bibnamefont
  {Rubio}},\ }\bibfield  {title} {\bibinfo {title} {Cavity
  quantum-electrodynamical polaritonically enhanced electron-phonon coupling
  and its influence on superconductivity},\ }\href
  {https://doi.org/10.1126/sciadv.aau6969} {\bibfield  {journal} {\bibinfo
  {journal} {Science Advances}\ }\textbf {\bibinfo {volume} {4}},\ \bibinfo
  {pages} {eaau6969} (\bibinfo {year} {2018})}\BibitemShut {NoStop}%
\bibitem [{\citenamefont {Dé}\ \emph {et~al.}(2022)\citenamefont {Dé},
  \citenamefont {Eckhardt}, \citenamefont {Kennes},\ and\ \citenamefont
  {Sentef}}]{LeDe_2022}%
  \BibitemOpen
  \bibfield  {author} {\bibinfo {author} {\bibfnamefont {B.~L.}\ \bibnamefont
  {Dé}}, \bibinfo {author} {\bibfnamefont {C.~J.}\ \bibnamefont {Eckhardt}},
  \bibinfo {author} {\bibfnamefont {D.~M.}\ \bibnamefont {Kennes}},\ and\
  \bibinfo {author} {\bibfnamefont {M.~A.}\ \bibnamefont {Sentef}},\ }\bibfield
   {title} {\bibinfo {title} {Cavity engineering of {H}ubbard ${U}$ via phonon
  polaritons},\ }\href {https://doi.org/10.1088/2515-7639/ac618e} {\bibfield
  {journal} {\bibinfo  {journal} {Journal of Physics: Materials}\ }\textbf
  {\bibinfo {volume} {5}},\ \bibinfo {pages} {024006} (\bibinfo {year}
  {2022})}\BibitemShut {NoStop}%
\end{thebibliography}%
	\onecolumngrid
	\clearpage
	
	{\allowdisplaybreaks
		
		\begin{center}
			\begin{center}
				\vspace{0.25cm}	\textbf{\large Supplemental Materials for ``Nonequilibrium phononic first-order phase transition in a driven fermion chain''}
			\end{center}
			\date{\today}
		\end{center}
			\maketitle
			{\allowdisplaybreaks
				\setcounter{equation}{0}
				
				\setcounter{equation}{0}
				\renewcommand{\theequation}{S\arabic{equation}}
				\setcounter{figure}{0}
				\renewcommand{\thefigure}{S\arabic{figure}}
				\setcounter{section}{0}
				\renewcommand{\thesection}{S\arabic{section}}
				\setcounter{table}{0}
				\renewcommand{\thetable}{S\arabic{table}}

				\section{S1. Hamiltonian model in $k$-space}
				In this section, we transfer the model in Eq.~(1) into the reciprocal space. We apply the following Fourier transformations for both electronic~(characterized by mode $k$) and phononic~(characterized by mode $q$) sectors, $c^\dagger_\ell = {} L^{1/2}\sum_k \exp(-i k \ell) c^\dagger_k$ and $a^\dagger_\ell = {} L^{1/2} \sum_q \exp(-i q \ell) a^\dagger_q$. With this, the electronic dispersion is given by $\omega_k = - 2 t_0 \cos(k)$. As the first approximation, we consider the infrared active phonons as well as we only take into account the dispersionless $q = 0$ phonon mode. The physical reason behind the zero-momentum phonon mode approximation can be understood from the fact that the atoms generating the phonons in a lattice are of the order of Angstroms, while the laser wavelengths are of the order of nanometers. From this separation of scales, the zero phonon mode dominates. It can also be justified by the fact that in most materials the average phonon frequency is much larger than the phonon bandwidth. Thus, the total Hamiltonian reads\begin{equation}\label{eq_s1}
					\mathcal{H} = {} \sum_k \omega_k c^\dagger_k c_k  +  \omega_0\, a^\dagger_0 a_0 + \red{\Big[\frac{g_{\rm l}}{\sqrt{L}} \big(a^\dagger_0 + a_0\big) + \frac{g_{\rm q}}{L} \big(a^\dagger_0 + a_0\big)^2\Big] }  \sum_k  \big(c^\dagger_k c_k - \langle c^\dagger_k c_k\rangle_{\rm eq}\big) + \mathcal{E}(t) \sqrt{L} \big(a^\dagger_0 + a_0\big)\, ,
				\end{equation}\red{where $g_{\rm l}$ describes the linear electron-phonon coupling}, $k$ belongs to the Brillouin zone from $-\pi$ to $\pi$ for a lattice constant $l_0 = 1$. Moreover, $L^{-1} \sum_k \langle c^\dagger_k c_k\rangle_{\rm eq} = 1/2$ is the equilibrium value of electron density. 
				
				\section{S2. Physical observables and quantum Master Equations}
				In this section, we define the physical observables and the corresponding expectation values for both electronic and phononic sectors as
				\begin{subequations}\label{eq_s2}
					\begin{align}
						&\hspace*{-0.15cm}q_{\rm ph}(t) =   \langle \frac{1}{\sqrt{L}} (a^\dagger_0  + a_0) \rangle (t):= \text{phonon displacement}\, ,\label{eq_s2a}\\
						&\hspace*{-0.15cm}p_{\rm ph}(t) =  \langle \frac{i}{\sqrt{L}} (a^\dagger_0  - a_0)\rangle (t):= \text{phonon momentum}\,, \label{eq_s2b}\\
						&\hspace*{-0.15cm}n_{\rm ph}(t) =  \langle \frac{1}{L} a^\dagger_0 a_0 \rangle (t):= \text{phonon occupation}\, ,\label{eq_s2c}\\
						&\hspace*{-0.15cm}\mathcal{Q}_{\rm ph}(t) =  \langle \frac{1}{L} (a^\dagger_0a^\dagger_0 + a_0 a_0) \rangle (t) :=  \text{squeezed-phonon displacement} ,\label{eq_s2d}\\
						&\hspace*{-0.15cm}\mathcal{P}_{\rm ph}(t) =  \langle \frac{i}{L} (a^\dagger_0a^\dagger_0 - a_0 a_0) \rangle (t) := \text{squeezed-phonon momentum} ,\label{eq_s2e}\\
						&\hspace*{-0.15cm}n_{{\rm e},k}(t) =  \langle c^\dagger_k c_k \rangle (t):= \text{$k$-component of electron number}\, .\label{eq_s2f}
					\end{align}
				\end{subequations}
				Before turning to the ordinary differential equations of motion~(EoM), it is also useful to define the electron density
				\begin{equation}\label{eq_s3}
					n_{\rm e}(t) = \frac{1}{L} \sum_k n_{{\rm e},k}(t)\, 
				\end{equation}
				for which the constraint $n_{\rm e}(t) \leq 1/2$ should be hold in the presence of quadratic electron-phonon coupling~(QEPC) $g_{\rm q}$. We would also mention that Lindemann's criterion~\cite{Lindemann1910} provides an estimate for the phonon occupation $n_{\rm ph}(t) \lessapprox 1$ to avoid lattice melting in a solid state system. This is also satisfied in our simulations.
				
				Now we take the role of dampings into account. To do so, we use the adjoint quantum master equation for an arbitrary observable $O(t)$ in an open system for which the Heisenberg equation of motion is extended to deduce the time evolution of the entire system~\cite{breuer2007theory,lindblad1976}. The equation describing the coherent evolution and the dissipator reads 
				\begin{equation}\label{eq_s4} 
					\langle \dot{O}\rangle (t) = {}   i\langle[\mathcal{H},O(t)]\rangle  + \frac{1}{2}\sum_\ell \gamma_\ell \Big<\big[\mathcal{L}_\ell^{\dagger},O(t)\big]\mathcal{L}_\ell  + \mathcal{L}_\ell^{\dagger}\big[O(t),\mathcal{L}_\ell\big]\Big> \, ,
				\end{equation}
				where the summation $\ell$ runs over all possible states in the Hilbert space and ${\mathcal{L}}_\ell$ are the time-independent Lindblad jump operators in the reduced system's Liouville space. The damping parameters are $\gamma_{\rm ph}$ and $\gamma_{\rm e}$, respectively, for the phonon and electron.
				
				The Heisenberg term, i.e. the first term of the above equation, is straightforward to be obtained for defined observables in Eq.~\eqref{eq_s2}. Turning to the second term of Eq.~\eqref{eq_s4}, the possible jump operators for the phononic sector can be each of $\mathcal{L}_\ell= a_0$, $a^\dagger_0$, $a_0 a^\dagger_0$, $a^\dagger_0 a_0$, $a_0 a_0$, and $a^\dagger_0 a^\dagger_0$. Let us label the corresponding damping parameters with $\gamma_{1}, \gamma_{2}, \gamma_{3}, \gamma_{4}, \gamma_{5}$,  and $\gamma_{6}$, respectively. Since the environment is supposed to be the same for all these channels, one can rewrite the above damping parameters as $\gamma_{\rm ph} r_{1}, \gamma_{\rm ph} r_{2}, \gamma_{\rm ph} r_{3}, \gamma_{\rm ph} r_{4}, \gamma_{\rm ph} r_{5},$ and $\gamma_{\rm ph} r_{6}$, respectively, in which $r_{\{1,\cdots,6\}}$ refer to the arrival rates of states after interacting with the phononic bath. For the two first operators, since the weak interaction with many atoms is supposed to simulate the coupling to a thermal bath~(Markovian approximation) with temperature $T$, the rates $r_{1}$ and $r_{2}$ must be related to each other by a Maxwell-Boltzmann factor, i.e.
				\begin{equation}\label{eq_s5}
					\frac{r_2}{r_1} = {} e^{-\widetilde{\omega}_0(t)/k_{\rm B}T} = {} \frac{\mathcal{N}_0(t)}{1+\mathcal{N}_0(t)}\, ,
				\end{equation}
				where $\mathcal{N}_0(t)$ is the mean number of energy quanta at time $t$ in the phonon mode corresponding to the modulated frequency $\widetilde{\omega}_0(t) = \omega_0 \sqrt{1+4 g_{\rm q} \Delta n_{\rm e}(t)/\omega_0}$, where $\Delta n_{\rm e}(t) = n_{\rm e}(t)-1/2$~\cite{PhysRevA.98.052129}. As for the cases $a^\dagger_0 a_0$ and $a_0 a^\dagger_0$, the corresponding states have the same arrival rates because they are not acting as ladder operators and do not change the states when interacting, thus, $r_4/r_3 = 1$. For the two last cases, however, we neglect order operators higher than bilinear and set their contribution to zero in the dissipation process, i.e. $r_5 = r_6 = 0$. Considering Eq.~\eqref{eq_s4}, after pretty straightforward calculations, the only Lindblad operators contributing to the dissipation effects originate from the non-conserving phonon operators, $a_0$ and $a^\dagger_0$, described by Eq.~\eqref{eq_s5}. 
				
				Finally, in order to make these coupled equations solvable, we use a mean-field approximation in which entanglement between electronic and phononic degrees of freedom are neglected. With these approximations, we then obtain the time evolution of the phononic observables as
				\begin{subequations} \label{eq_s6} 
					\begin{align}  
						\dot{q}_{\rm ph}(t) = {} & + \omega_0 \,p_{\rm ph}(t) -  \frac{\gamma_{\rm ph}}{2} [1+2\mathcal{N}_0(t)]q_{\rm ph}(t)\, ,\label{eq_s6a}\\ 
						\dot{p}_{\rm ph}(t) =  {} &- \big[\omega_0 + 4 g_{\rm q} \Delta n_{\rm e}(t)\big]q_{\rm ph}(t)  \hspace*{-0.05cm}-\hspace*{-0.05cm}2 \red{\big(\mathcal{E}(t) + g_{\rm l}\, \Delta n_{\rm e}(t)\big)} - \frac{\gamma_{\rm ph}}{2} [1+2\mathcal{N}_0(t)]p_{\rm ph}(t) ,\label{eq_s6b}\\
						\dot{n}_{\rm ph}(t) = {} & -\red{\big(\mathcal{E}(t) + g_{\rm l}\, \Delta n_{\rm e}(t)\big)}\, p_{\rm ph}(t) - 2 g_{\rm q} \Delta n_{\rm e}(t) \mathcal{P}_{\rm ph}(t)- \gamma_{\rm ph} [n_{\rm ph}(t)-\mathcal{N}_0(t)] \, ,\label{eq_s6c}\\
						\dot{\mathcal{Q}}_{\rm ph}(t) = {} & + 2 \big[\omega_0 + 2 g_{\rm q} \Delta n_{\rm e}(t)\big] \mathcal{P}_{\rm ph}(t) + 2 \red{\big(\mathcal{E}(t) + g_{\rm l}\, \Delta n_{\rm e}(t)\big)} \, p_{\rm ph}(t)  - \gamma_{\rm ph} [\mathcal{Q}_{\rm ph}(t)-\mathcal{N}_0(t)]\,, \label{eq_s6d}\\
						\dot{\mathcal{P}}_{\rm ph}(t) = {}& -2 \big[\omega_0 + 2 g_{\rm q} \Delta n_{\rm e}(t)\big] \mathcal{Q}_{\rm ph}(t)  -2 \,\red{\big(\mathcal{E}(t) + g_{\rm l}\, \Delta n_{\rm e}(t)\big)}\,q_{\rm ph}(t)  -4 g_{\rm q} \big[2 n_{\rm ph}(t) + \rfrac{1}{L}\big]\Delta n_{\rm e}(t)-\gamma_{\rm ph} [\mathcal{P}_{\rm ph}(t)-\mathcal{N}_0(t)]\, .\label{eq_s6e}
					\end{align} 
				\end{subequations}As the laser drive is what takes us out of equilibrium, there is no displacement/movement for the phononic sector in equilibrium, leading to the initial conditions $O_{\rm ph}(0) = 0$.
				
				For the single electron $O(t) = n_{{\rm e},k}(t)$, we again consider all possible Lindblad operators $c_k$, $c^\dagger_k$, $c_k c^\dagger_k$, and $c^\dagger_k c_k$. Again, we do not consider higher order terms including electron-electron interaction and would obtain bilinear EoMs. Then, the arrival rates of states to the electron after interacting with the phononic bath follow the same argument as
				\begin{equation}\label{eq_s7}
					\frac{r_2}{r_1} = {} e^{-\widetilde{\omega}_k(t)/k_{\rm B}T} = {} \frac{\mathcal{N}_k(t)}{1-\mathcal{N}_k(t)}\, ,
				\end{equation}
				where $\mathcal{N}_k(t)$ is the mean number of energy quanta at time $t$ in the electron mode corresponding to the modulated dispersion \red{$\widetilde{\omega}_k(t) = \omega_k + g_{\rm l} q_{\rm ph}(t)+ g_{\rm q} q^2_{\rm ph}(t)$}~\cite{PhysRevA.98.052129}. The fact that QEPC acts as a chemical potential in the electronic dispersion stems from the fact that the phonon is coupled to the local electron number in our model, leading to a shift in the electron energy. \red{While the LEPC only makes the dispersion oscillatory around its equilibrium value with no shift.}
				
				Within the same manner, the electron conserving jump operators $c_k c^\dagger_k$, $c^\dagger_k c_k$ lead to zero contributions to the dynamical properties of electron~($r_4/r_3 = 1$).  Ultimately, the time evolution of the electron number is obtained as
				\begin{equation}\label{eq_s8}
					\dot{n}_{{\rm e},k}(t) = {} - \gamma_{\rm e} [n_{{\rm e},k}(t)-\mathcal{N}_k(t)] \,.
				\end{equation}
				
				As an extension to the model, placing material into an optical cavity can be considered to circumvent detrimental heating, which is a fundamental problem for driving materials with classical light. This, in turn, enhances the light-matter coupling for which a few photon states of the cavity can influence the matter’s degrees of freedom~\cite{Schlawin2022}. Generally, to treat both light and matter quantum mechanically, one has to deal with host hybrid states as a mixture of both light and matter degrees of freedom. For instance, it has been proposed that coupling cavity modes to the phonons involved in electronic pairing~\cite{doi:10.1126/sciadv.aau6969} or by directly coupling to the electronic degrees of freedom, the superconductivity in a cavity can be influenced~\cite{LeDe_2022}.
				
				\section{S3. Nonequilibrium steady state}
				\renewcommand{\thefigure}{S2-1}
				\begin{figure}[t]
					\centering
					\includegraphics[width=0.4\linewidth]{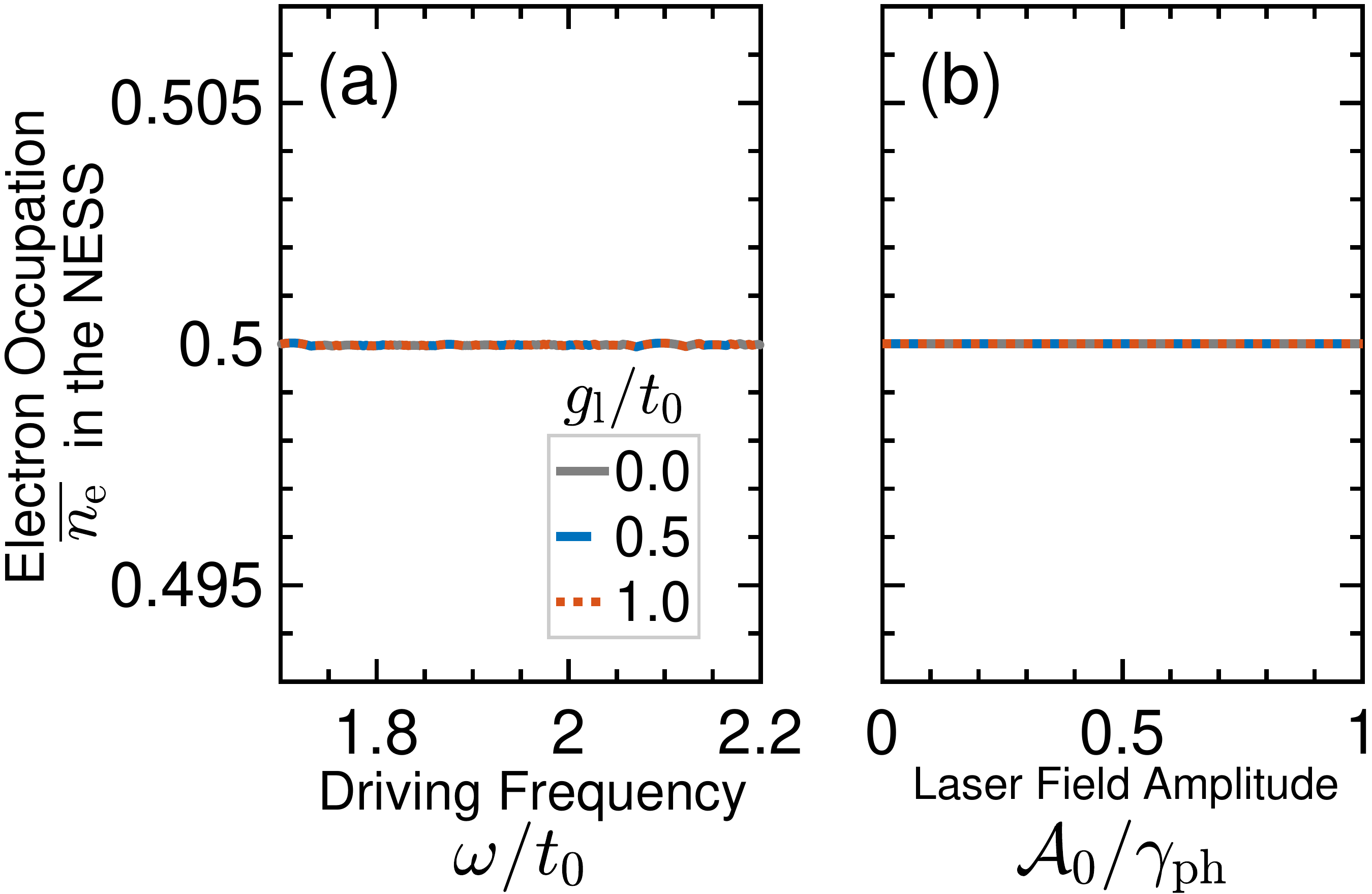}
					\caption{\red{Dressed electron occupation for various LEPCs as a function (a) of driving frequency at $\mathcal{A}_0 = 0.038$ THz and (b) of laser field amplitude at $\omega = 4.44$ THz. Fixed parameters are phonon frequency $\omega_0 = 4.8$, $\gamma_{\rm ph} = 0.24$, and $\gamma_{\rm e} = 0.0024$~(all in units of THz). Negligible effect of LEPC is evident.}} 
					\label{fS31_SM}
				\end{figure}
				
				\red{To reveal the local depopulation of electrons due to EPC, we calculate electron occupation in the nonequilibrium steady state~(NESS). At $T = 0$, electron dynamics depend on the mode $k$ given by $n_{{\rm e},k}(t) = \langle  c^\dagger_k c_k\rangle (t) = - \gamma_{\rm e} \int_0^t \big(n_{{\rm e},k}(t')-\mathcal{N}_k(t')\big) {\rm d} t'$ with time-dependent Fermi-Dirac mean number $\mathcal{N}_k(t) = 1/[e^{\widetilde{\omega}_k(t)/k_{\rm B}T}+1]$ of energy quanta $\widetilde{\omega}_k(t) = \omega_k + g_{\rm l} q_{\rm ph}(t) + g_{\rm q} q^2_{\rm ph}(t)$; $\omega_k = -2 t_0 \cos(k)$ and $q_{\rm ph}(t) =  \langle \frac{1}{\sqrt{L}} (a^\dagger_0  + a_0) \rangle (t)$, respectively, denote the bare electronic dispersion and the phonon displacement. The time-dependency of electron number arises from the self-consistent alteration of the occupations. Time-dependent electronic dispersion $\widetilde{\omega}_k(t)$ reflects the mutual feedback through coupling in $\mathcal{N}_k(t)$ which is simply a step function~\cite{PhysRevA.98.052129,lindblad1976}. Thus, we have ${L}^{-1} \sum_k \mathcal{N}_k(t) = \tilde{k}_{\rm F}(t)/\pi$, where $\tilde{k}_{\rm F}(t) = \pi/2 - [g_{\rm l} q_{\rm ph}(t)+g_{\rm q} q^2_{\rm ph}(t)]/2 t_0$ is obtained via linearization of bare electronic dispersion around the Fermi level. Accordingly, we obtain electron occupation from squeezing of the phonon state via
					\begin{equation}\label{eq_3}
						n^{\rm NESS}_{\rm e}(t) = \frac{1}{2} - \frac{g_{\rm l} q_{\rm ph}(t)+g_{\rm q} q^2_{\rm ph}(t)}{2 \pi t_0}\, .
					\end{equation} Averaging over oscillations in the NESS leads to $\overline{n_{\rm e}} = 1/2 - g_{\rm q} \overline{q^2_{\rm ph}}/2 \pi t_0$ because $\overline{q_{\rm ph}} = 0$ such that only QEPC contributes effectively to the NESS electron response, see Fig.~\ref{fS31_SM}. From these points, we neglect the LEPC term in what follows.}
				
				To establish the NESS by steady laser driving, we first investigate the evolution of the resonance~($\omega = \omega_0$) phononic and electronic responses upon tuning the QEPC across the nonequilibrium, as shown in Fig.~\ref{fS21_SM}. Due to the weak coupling of a single Einstein phonon to an ensemble of phonons~(bath), we set phonon damping rate $\gamma_{\rm ph}$ to a value of order 5\% of the phonon energy, consistent with the experimental recombination rates of certain quasi-1D Mott insulators~\cite{PhysRevLett.112.117801}. Further, the electron mode is in general damped to a phononic bath weaker than Einstein phonon, i.e., $\gamma_{\rm e} < \gamma_{\rm ph}$, and accordingly we set $\gamma_{\rm e}$ to a value of order 0.1\% of the electron hopping energy. 
				
				Phonon occupation pumps into a steady state exponentially due to the laser field and reaches the NESS at approximately four-time constants of the phonon system $2/\gamma_{\rm ph}$. From the fact that the power of the laser is proportional to the squared amplitude of the laser field, the plateau value of decoupled phase $g_{\rm q} = 0$ reaches $(\mathcal{A}_0/\gamma_{\rm ph})^2$ similar to a damped driven harmonic oscillator, in agreement with Ref.~\cite{yarmohammadi2020dynamical}. For $g_{\rm q}=0$, the electron number displays the equilibrium plateau 1/2. For $g_{\rm q} \neq 0$, a NESS forms with deviations in phononic and electronic plateaus stemming from the feedback between nonequilibrium phonons and dressed electronic dispersion $\widetilde{\omega}_k(t) = \omega_k + g_{\rm q} q^2_{\rm ph}(t)$. Subtracting averages from the time evolutions in the NESS gives rise to trackable oscillations, as shown in inset panels of Figs.~\ref{fS21_SM}(a) and~\ref{fS21_SM}(b). The point refers to the oscillation frequency; both quantities oscillate with the frequency $2 \,\omega_0$ -- frequency doubling effect -- in the NESS accompanied by a phase shift $\varphi$ due to QEPC. As an explanation, the decrease~(increase) in the amplitude of NESS for the phonon~(electron) sector with QEPC is a standard consequence of energy level repulsion. 
				\renewcommand{\thefigure}{S2-2}
				\begin{figure}[t]
					\centering
					\includegraphics[width=0.75\linewidth]{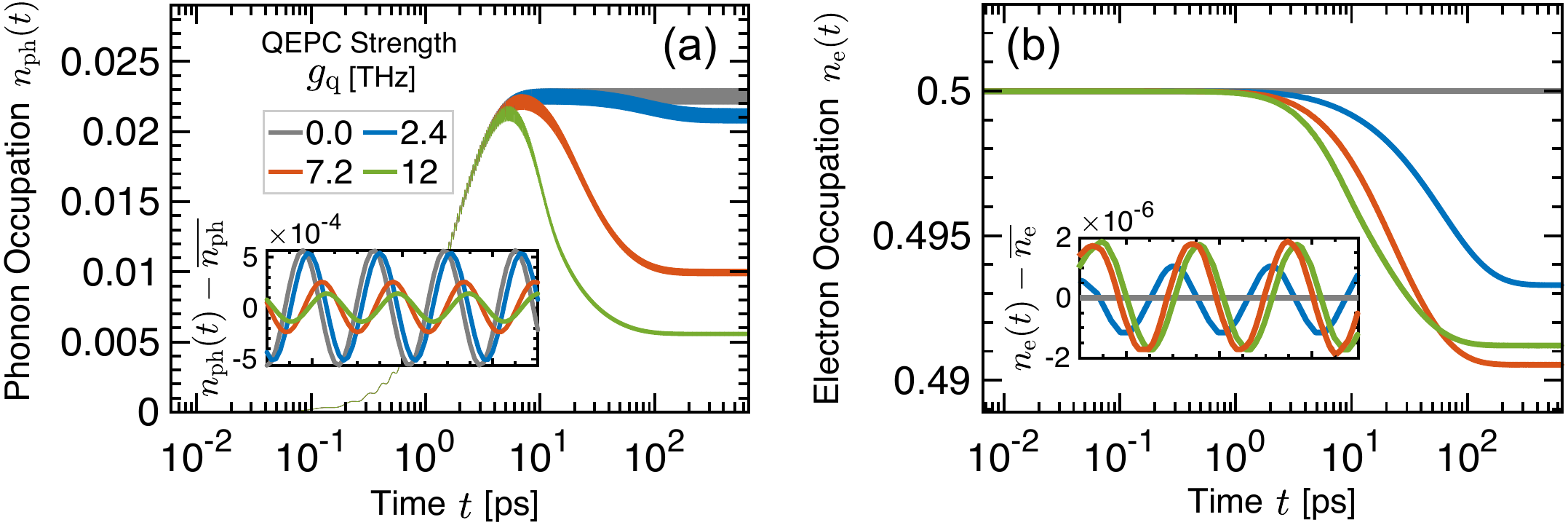}
					\caption{Time evolution of dressed (a) phonon and (b) electron occupations for various QEPCs at $\mathcal{A}_0 = 0.038$, resonant $\omega = \omega_0 = 4.8$, $\gamma_{\rm ph} = 0.24$, and $\gamma_{\rm e} = 0.0024$~(all in units of THz). Both sectors show a frequency doubling effect (inset panels) in the NESS accompanied by a phase shift $\varphi$ due to QEPC.} 
					\label{fS21_SM}
				\end{figure}
				
				\red{It is worth mentioning that our dissipation model provides an advantage compared to other studies~\cite{Sous2021,Kennes2017} for the stability of QEPC model. The renormalized phonon  frequency in the NESS $\overline{\widetilde{\omega}_0}  = \omega_0 \sqrt{1+4 g_{\rm q} [\overline{n_{\rm e}} - 1/2]/\omega_0}$ is linked to the renormalized oscillator stiffness $\overline{\widetilde{\mathcal{K}}} = \overline{\widetilde{\omega}_0}^2$ setting the oscillator mass $M = 1$. Thus, demanding a positive stiffness $\overline{\widetilde{\mathcal{K}}} > 0$ and considering our mean-field-type approximation, which replaces the quantized on-site electron occupation by an average value that only weakly~(maximum 2\% or 3\%) deviates from $1/2$, we find the stability condition of the QEPC model for $g_{\rm q}/\omega_0 < 25$ which is always achievable for all materials. Thereby, no instability occurs.}
				
				\section{S4. Energy flows through the electron-phonon system}
				
				In this section, we intend to focus on the energies per unit of time in the model. For a better understanding of the presented energy flows in the following, we draw Fig.~\ref{fS4}(a) to track the context. The input energy $\mathcal{J}^{\rm l \to p}$ is due to the laser field, which goes into the driven phonon. The driven phonon splits this input energy into the energy entering the QEPC part and the phononic bath, given respectively by $\mathcal{J}^{\rm p \to QEPC}$ and $\mathcal{J}^{\rm p \to b}$. The energy entering the QEPC part is again divided into two parts $\mathcal{J}^{\rm QEPC \to e}$ and $\mathcal{J}^{\rm QEPC \to b}$ referring, respectively, to the energy entering the electron and the bath. Finally, the electron induces the energy flow of $\mathcal{J}^{\rm e \to b}$ to the bath through the dissipation. These powers can directly be obtained from the EoMs of phonon occupation and electron number in Eqs.~\eqref{eq_s6c} and~\eqref{eq_s8}:
				\begin{subequations}\label{eq_s14}
					\begin{align}
						\mathcal{J}^{\rm l \to p}(t) = {} & -\mathcal{E}(t) \omega_0\, p_{\rm ph}(t)\, ,\label{eq_17a}\\
						\mathcal{J}^{\rm p \to QEPC}(t) = {} & +2g_{\rm q}\, \omega_0\, \Delta n_{\rm e}(t)\, \mathcal{P}_{\rm ph}(t)\, ,\label{eq_17b}\\
						\mathcal{J}^{\rm p \to b}(t) = {} & +\gamma_{\rm ph}\,\omega_0 \, n_{\rm ph}(t)\, ,\label{eq_17c}\\
						\mathcal{J}^{\rm QEPC \to b}(t) = {} & + g_{\rm q}\, \left[\gamma_{\rm e} +\gamma_{\rm ph} \right]\,\Delta n_{\rm e}(t)\,q^2_{\rm ph}(t) \, ,\label{eq_17d}\\
						\mathcal{J}^{\rm QEPC \to e}(t) = {} & +g_{\rm q}\, q^2_{\rm ph}(t) \gamma_{\rm e}\Big[\frac{1}{2} - \frac{1}{L} \sum_k \mathcal{N}_k(t)\Big]\,,\\
						\mathcal{J}^{\rm e \to b}(t) = {} & + \frac{\gamma_{\rm e}}{L}\,\sum_k \omega_k\, \big[n_{{\rm e},k}(t)-\mathcal{N}_k(t)\big]\, .\label{eq_17f}
					\end{align}
				\end{subequations}Although most of the above expressions are clear from the EoMs, Eq.~\eqref{eq_17d} needs more clarification: $J^{\rm QEPC \to b}(t)$ is obtained using the time-derivative of QEPC Hamiltonian in Eq.~\eqref{eq_s1} via
				\begin{equation}
					\frac{1}{L} \partial_t \langle \mathcal{H}_{\rm e-ph}\rangle(t) = {}  \partial_t [g_{\rm q}\,q^2_{\rm ph}(t)\,\Delta n_{\rm e}(t)] = {}  \mathcal{J}^{\rm p \to QEPC}(t) - \left[\mathcal{J}^{\rm QEPC \to b}(t)+\mathcal{J}^{\rm QEPC \to e}(t)\right]\, ,
				\end{equation}
				which the very small energy entering the electron from the QEPC part is stemming from the reflection of the dressed electronic dispersion from the bath to the QEPC subsystem. 
				\renewcommand{\thefigure}{S4}
				\begin{figure}[t]
					\centering
					\includegraphics[width=0.75\linewidth]{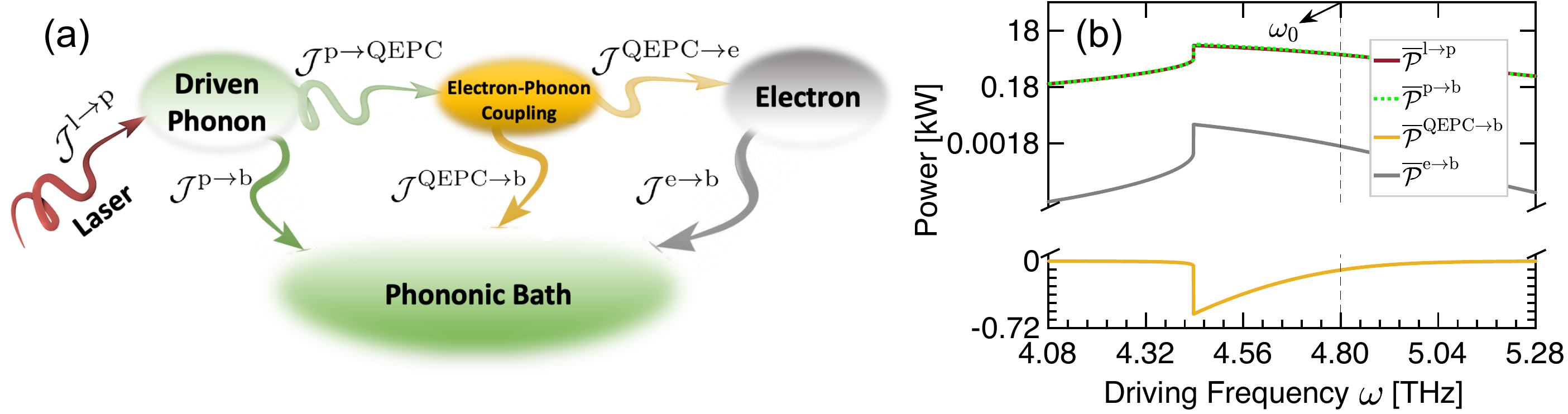}
					\caption{(a) Schematic picture of the energy flows within a driven-dissipative fermionic chain; $\mathcal{J}^{\rm l \to p}$ is the uptake of laser energy by the driven phonon, $\mathcal{J}^{\rm p \to QEPC}$ is energy flowing out of the driven phonon due to the presence of the QEPC, and $\mathcal{J}^{\rm p \to b}$ is the energy flowing from the driven phonon and directly to the phononic bath. Furthermore, $\mathcal{J}^{\rm QEPC \to e}$ and $\mathcal{J}^{\rm QEPC \to b}$ are, respectively, the energy flowing out of the QEPC part to the electron and bath. Finally, the electron induces the energy flow of $\mathcal{J}^{\rm e \to b}$ to the bath through the dissipation. (b) Input and output powers through the electron-phonon system with the same parameters as Fig.~2(a) and Fig.~2(b) for $g_{\rm q}/\omega_0 = 2$. Total power sum rule $\overline{\mathcal{J}}^{\rm l \to p} = \overline{\mathcal{J}}^{\rm p \to b} + \overline{\mathcal{J}}^{\rm QEPC \to b} + \overline{\mathcal{J}}^{\rm e \to b}$ is evident from the driving to the final stage of dissipation.}
					\label{fS4}
				\end{figure}
				
				Having the explicit form of energy flows, we can explore the physical sum rules which should be valid at each stage. It should be noted that, after taking the average of the late-time signals over one period, the input energy from the laser driving field should satisfy $\overline{\mathcal{J}}^{\rm l \to p} = \overline{\mathcal{J}}^{\rm p \to b} + \overline{\mathcal{J}}^{\rm QEPC \to b} + \overline{\mathcal{J}}^{\rm e \to b}$ in the NESS, originating from the energy conservation. In addition to this general sum rule, three further rules are valid for the subprocesses: (i) $\overline{\mathcal{J}}^{\rm l \to p} = \overline{\mathcal{J}}^{\rm p \to QEPC} + \overline{\mathcal{J}}^{\rm p \to b}$, (ii) $\overline{\mathcal{J}}^{\rm p \to QEPC} = \overline{\mathcal{J}}^{\rm QEPC \to e} + \overline{\mathcal{J}}^{\rm QEPC \to b}$ and (iii) $\overline{\mathcal{J}}^{\rm QEPC \to e} = \overline{\mathcal{J}}^{\rm e \to b}$.
				
				Finally, with the help of energy flows, one can use a sample with a thickness $w$, area $a$, and the molar density $\rho$ to obtain the power in kW as
				\begin{equation}
					\mathcal{P}^{\mcirc \to \msquare}(t) = w a \rho\,\mathcal{J}^{\mcirc \to \msquare}(t)\, .
				\end{equation}
				The input energy fluence $\mathcal{P}^{\rm l \to p}$ due to the laser field excites the phonon and it is an absorbance spectrum proportional to the phonon occupation, as shown in Fig.~\ref{fS4}(b). The net energy flow bears a close resemblance to Figs.~\ref{fS31_SM}(a) and~\ref{fS31_SM}(b) with additional weighting factors of $\omega_0$ and $\omega_k$~(see Eq.~\ref{eq_s14} of the SM). It can be seen that the majority of the laser energy flows directly to the phononic bath and a relatively tiny fraction of it~(negligible) can be absorbed by the electron. Moreover, the negative sign of $ \overline{\mathcal{P}}^{\rm QEPC \to b}$ implies a small energy flow from the bath due to the QEPC term. Therefore such an absorbance can be used for quantitative analysis. In Fig.~\ref{fS4}(b), one may compute the net power in the NESS, $\overline{\mathcal{P}}^{\rm p \to b} + \overline{\mathcal{P}}^{\rm QEPC \to b} + \overline{\mathcal{P}}^{\rm e \to b}$, which by the sum rules at each step matches $\overline{\mathcal{P}}^{\rm l \to p}$ for energy conservation.
				
				\section{S5. Effective force treatment of the nonequilibrium phase transition}
				
				As the input laser drive is a continuous wave field, the output coherent oscillations in the NESS must synchronize with it. To understand the phase shift caused by QEPC, we Fourier decompose the phonon oscillations in the NESS, $q^{\rm NESS}_{\rm ph}(t) = \sum_n q_n e^{i n \omega\,t}$ and substitute it in Eqs.~\eqref{eq_s6a} and~\eqref{eq_s6b}, leading to~(at $T = 0$, $\mathcal{N}_0(t) \to  0$)
				\begin{subequations}
					\begin{align}
						i n \omega q_n = {} & \omega_0 p_n - \frac{\gamma_{\rm ph}}{2} q_n\, ,\\
						i n \omega p_n = {} & - \omega_0 q_n - 4 q_{\rm q} \sum_{n'} \Delta n_{{\rm e},n'} q_{n-n'} + 2 g_{\rm q} q_n - 2 \mathcal{A}_0 - \frac{\gamma_{\rm ph}}{2} p_n\, ,
					\end{align}
				\end{subequations}
				As seen in the numerics, the dominant harmonics are $n=1$ for $q_{\rm ph}$ and $n'=0$ for $n_{\rm e}$ (henceforth, we use the notation $n_{{\rm e},0} = \overline{n_{\rm e}}$ since the zeroth Fourier harmonics is that of the averaged plateau in the NESS):
				\begin{subequations}\label{eq_S10}
					\begin{align}
						q_1 = {} &\frac{2 \mathcal{A}_0 \omega_0}{\omega^2 - \omega^2_0 + \frac{\gamma^2_{\rm ph}}{4} - 4g_{\rm q} \omega_0(\overline{n_{\rm e}} - 1/2) - i \gamma_{\rm ph} \omega} = |q_1| e^{i \varphi}\, ,\label{eq_S10a}\\
						|q_1| = {} &\frac{2 \mathcal{A}_0 \omega_0}{\sqrt{[\omega^2 - \omega^2_0 + \frac{\gamma^2_{\rm ph}}{4} - 4g_{\rm q} \omega_0(\overline{n_{\rm e}} - 1/2)]^2 + \gamma^2_{\rm ph} \omega^2}}\, , \label{eq_S10b}\\
						\varphi = {} &\arctan\Bigg(\frac{\gamma_{\rm ph} \omega}{\omega^2 - \omega^2_0 + \frac{\gamma^2_{\rm ph}}{4} - 4g_{\rm q} \omega_0(\overline{n_{\rm e}} - 1/2)}\Bigg)\, .\label{eq_S10c}
					\end{align}
				\end{subequations}\renewcommand{\thefigure}{S5-1}\begin{figure}[t]
					\centering
					\includegraphics[width=0.75\linewidth]{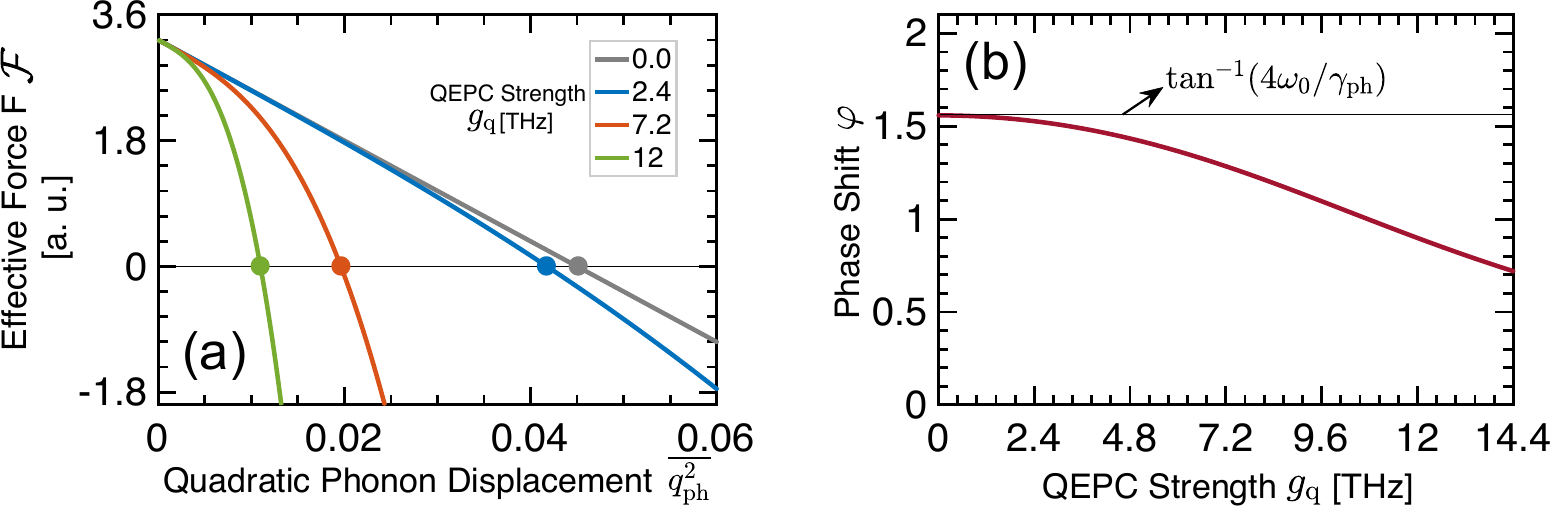}
					\caption{(a) Effective force $\mathcal{F}(\mathcal{X})$, Eq~\eqref{eq_S11a}, with $\mathcal{X} = \overline{q^2_{\rm ph}}$ for various QEPCs at $\mathcal{A}_0 = 0.038$, resonant $\omega = \omega_0 = 4.8$, $\gamma_{\rm ph} = 0.24$, and $\gamma_{\rm e} = 0.0024$~(all in units of THz). (b) Deviation from the decoupled phase shift, i.e. $\arctan(4\omega_0/\gamma_{\rm ph})$ observed in inset Figs.~\ref{fS31_SM}(a) and~\ref{fS31_SM}(b) in the NESS, is due to QEPC at $\mathcal{X} \simeq 0.0197$~(see the orange line in (a)).} 
					\label{fS3_SM}
				\end{figure}From a physical point of view, the driven phonon typically exhibits different characteristic behaviors far from equilibrium compared to equilibrium that are not generally described by a minimization principle. As we deal with a damped driven dressed (by the electron) harmonic oscillator, there is an effective force acting on the phonon and microscopically on $|q_1|$ far from equilibrium. This force can be obtained from setting $\ddot{q}_{\rm ph} = \dot{q}_{\rm ph} = 0$ in the NESS. Using the relations $\overline{q^2_{\rm ph}} = |q_1|^2/2$ and $\overline{n_{\rm e}} = 1/2 - g_{\rm q} \overline{q^2_{\rm ph}}/2 \pi t_0$ stemming from the dressed Fermi momentum $\tilde{k}_{\rm F}$, Eq.~\eqref{eq_S10b} can be rewritten as a cubic equation~$\mathcal{F}(\mathcal{X}) = 0$ for $\mathcal{X}=\overline{q^2_{\rm ph}}$, satisfying the physics behind the relative effective force~(assuming atomic mass $= 1$) discussed above. Here we use this relative effective force to find $ |q_1|  = \sqrt{2 \overline{q^2_{\rm ph}}}$ required for the phase shift. Thus,
				\begin{subequations}\label{eq_S11}
					\begin{align}
						& \mathcal{F}(\mathcal{X}) = {} a \mathcal{X}^3 + b \mathcal{X}^2 + c\mathcal{X} + d \, ,\label{eq_S11a}\\
						& a = {} -4 \frac{g^4_{\rm q} \omega^2_0}{\pi^2 t^2_0} \quad , \quad b = {}  -4 \frac{g^2_{\rm q} \omega_0}{\pi t_0} \Big(\omega^2 - \omega^2_0 + \frac{\gamma^2_{\rm ph}}{4}\Big)\quad , \quad c = {} -\gamma^2_{\rm ph} \omega^2 - \Big(\omega^2 - \omega^2_0\Big)^2\quad , \quad 
						d = {}2 \mathcal{A}^2_0 \omega^2_0\, .\label{eq_S11b}
					\end{align}
				\end{subequations}
				As shown in Fig.~\ref{fS3_SM}(a), a single root appears for $\mathcal{F}(\mathcal{X})$ when the laser drive is in resonance with the phonon. This root should be plugged into $\overline{n_{\rm e}}$ to find the phase shift $\varphi$ at $\omega = \omega_0$, resulting in $\varphi(\mathcal{X}) =\arctan(\gamma_{\rm ph} \omega_0/[\gamma^2_{\rm ph}/4 + 2 g^2_{\rm q} \omega_0 \mathcal{X}/\pi t_0])$, as shown in Fig.~\ref{fS3_SM}(b). This phase shift decreases with QEPC as the quadratic phonon displacement is decreased.
				
				By analyzing $\mathcal{F}(\mathcal{X})$ diagram using Eq.~\eqref{eq_S11a} for off-resonance conditions, we can attribute the origin of phase transition to shifts of the cubic function, leading to creation/annihilation of roots. The off-resonance nonequilibrium dynamics of the spinless fermionic chain subject to a laser drive can broadly be divided into three regions depending on the QEPC strength, namely $\omega < \omega_{\rm c}$, $\omega = \omega_{\rm c}$, and $\omega > \omega_{\rm c}$, where $\omega_{\rm c}$ is the threshold driving frequency at which the phase transition takes place. For these regimes, we plot $\mathcal{F}(\mathcal{X})$ as a function of $\mathcal{X}$ in Fig.~\ref{fS4_SM}(a). The extrema of $\mathcal{F}(\mathcal{X})$ from Eq.~\eqref{eq_S11a} occur when ${\rm d} \mathcal{F}(\mathcal{X})/{\rm d}\mathcal{X} = 0$, so, two stationary points locate at $\mathcal{X}_1 = -(b/3a) +\sqrt{\Delta}/6a$ and $\mathcal{X}_2 = -(b/3a) -\sqrt{\Delta}/6a$, where $\Delta = 4b^2 - 12 a c$. From Eq.~\eqref{eq_S11b}, $a$ and $c$ coefficients are always positive, while $d$ is always negative. For $\omega < \omega_0$, $b$ becomes negative, while it is positive at $\omega = \omega_0$ and $\omega > \omega_0$. Here we address $\omega < \omega_0$. For various driving frequencies in the current set of parameters, there is a critical $\omega = \omega_{\rm c} = 4.44$ THz at which two of the three distinct roots become identical at ${\rm Re}[\mathcal{X}_1]$. For $\omega < \omega_{\rm c}$, the first and third roots are stable solutions, while the second root is unstable; it can be understood from the slopes around the roots. However, the first solution is the relevant physical one based on the nature of EoMs in our coupled model with a vacuum ground state. As an explanation, one would argue that the laser cannot strongly excite phonons at low frequencies~(before the strongest response at the phase transition point) and the weakest phonon displacement is expected to be detected, which is that of the first solution. To physically interpret the behaviors at $\omega = \omega_{\rm c}$, one would argue that the force acting on the phonon instantaneously vanishes and the pattern becomes static, hence, a sharp jump to the third root for restoring dynamic patterns happens. This is where the first-order phase transition emerges. For $\omega > \omega_{\rm c}$, the third root is the only left one as we approach the phonon frequency $\omega_0 = 4.8$ THz.
				\renewcommand{\thefigure}{S5-2}
				\begin{figure}[t]
					\centering
					\includegraphics[width=1\linewidth]{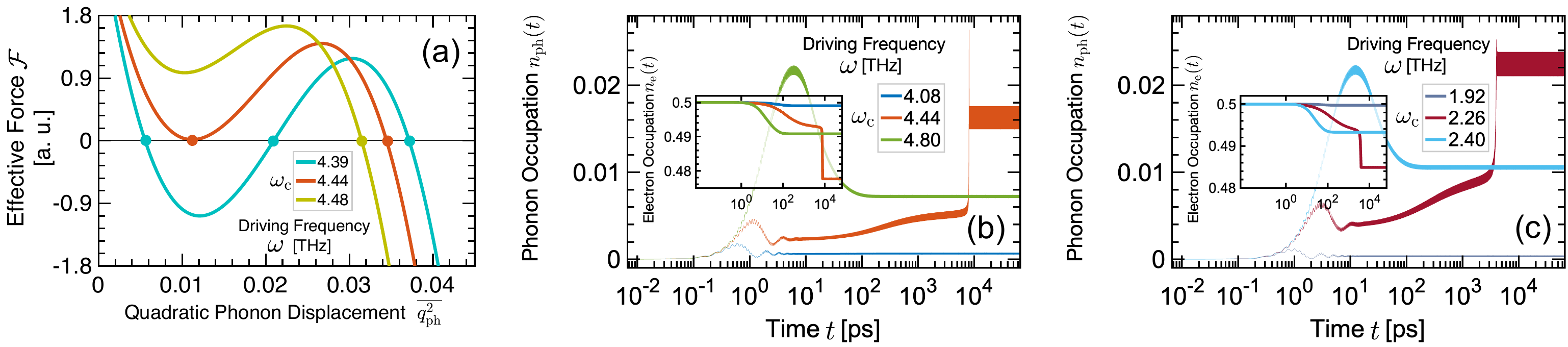}
					\caption{(a) Effective force, Eq~\eqref{eq_S11a}, characterized by $\mathcal{X} := \overline{q^2_{\rm ph}}$ for off-resonance driving frequencies with the phonon frequency $\omega_0 = 4.8$ THz. The threshold driving frequency at which the phase transition occurs is $\omega_{\rm c} = 4.44$ THz. Off-resonance time evolution of dressed phonon occupation~(main panels) and electron number~(inset panels) for phonon frequency (b) $\omega_0 = 4.8$ THz with $\omega_{\rm c} = 4.44$ THz and (c) $\omega_0 = 2.4$ THz with $\omega_{\rm c} = 2.26$ THz and three regimes of driving frequencies below, at, and above $\omega_{\rm c}$ at $g_{\rm q}/\omega_0 = 2$, $\mathcal{A}_0 = 0.038$, $\gamma_{\rm ph} = 0.24$, and $\gamma_{\rm e} = 0.0024$~(all in units of THz).} 
					\label{fS4_SM}
				\end{figure}
				
				To find the exact analytical expression for $\omega_{\rm c}$, we use the fact that, at $\omega = \omega_{\rm c}$, the imaginary part of $\mathcal{X}_1$ for $\Delta < 0$ vanishes. Thus, one immediately finds $\omega_{\mathcal{X}_1} = \sqrt{\omega_0^2 + \frac{\gamma_{\rm ph}^2}{2} - \sqrt{3} \gamma_{\rm ph} \omega_0}$ to satisfy $\Delta < 0$. Afterwards, we obtain $\mathcal{X}_1 = -\tilde{b}|_{\omega_{\mathcal{X}_1}}/3a$ with
				\begin{equation}
					\tilde{b} = {}  \frac{g^2_{\rm q} \omega_0 \gamma_{\rm ph}}{\pi t_0} \Big(3 \gamma_{\rm ph} - 4 \sqrt{3} \omega_0\Big)\, .
				\end{equation}
				Due to the weak coupling of a single Einstein phonon to an ensemble of phonons~(bath), $\gamma_{\rm ph} < \omega_0$ always holds, implying that the above characteristic parameter $\tilde{b}$ is physically always negative to have the expected positive $\mathcal{X}_1 = \overline{q^2_{\rm ph}}$. Eventually, plugging $\mathcal{X}_1 = -\tilde{b}/3a$ into Eq.~\eqref{eq_S11a} yields the following solution for $\omega_{\rm c}$
				\begin{equation}\label{eq_S13}
					\omega_{\rm c} = \sqrt{\omega^2_0 - \frac{2\sqrt{3}}{3} \gamma_{\rm ph} \omega_0 + \frac{1}{2} \sqrt{\gamma_{\rm ph}^4-\frac{\pi t_0\tilde{b}}{3 g^2_{\rm q}\omega_0} - 4 \gamma_{\rm ph}^2 \omega_0^2- \frac{96 g^4_{\rm q} \omega^4_0 \mathcal{A}^2_0}{\pi^2 t_0^2 \tilde{b}}}}\, .
				\end{equation}
				For the set of parameters $\omega_0 = 4.8$, $g_{\rm q} = 9.6$, $\mathcal{A}_0 = 0.038$, $\gamma_{\rm ph} = 0.24$, and $\gamma_{\rm e} = 0.0024$~(all in units of THz), one exactly obtains $\omega_{\rm c} = 4.44$ THz, in excellent agreement with above numerical findings.
				\renewcommand{\thefigure}{S5-3}
				\begin{figure}[b]
					\centering
					\includegraphics[width=0.75\linewidth]{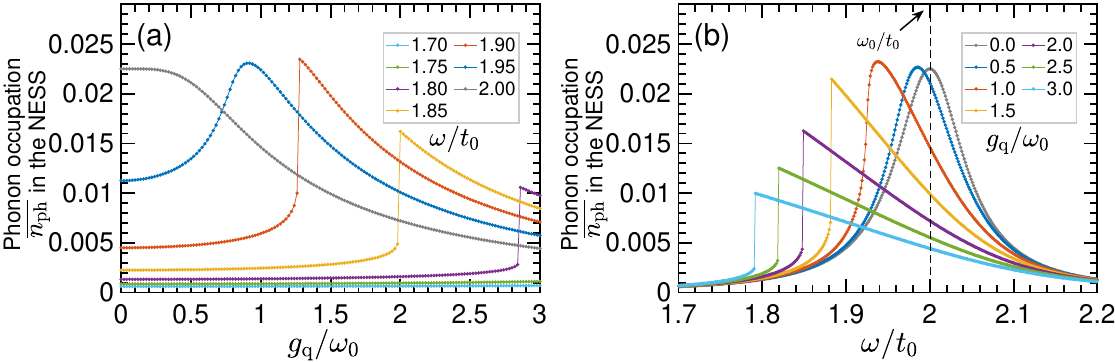}
					\caption{\red{Phonon occupation in the NESS as a function (a) of QEPC for various off-resonance driving frequencies and (b) of driving frequency for various QEPCs with the phonon frequency $\omega_0/t_0 = 2$. The critical QEPC and driving frequency at which the phase transition occurs is different for different set of parameters, as confirmed by Eq.~\eqref{eq_S13}. Other parameters are fixed at $\mathcal{A}_0 = 0.038$, $\gamma_{\rm ph} = 0.24$, and $\gamma_{\rm e} = 0.0024$~(all in units of THz)}.} 
					\label{fS53_SM}
				\end{figure}
				
				It should be noted that the Lindblad approach applies in the Markovian limit of a weak system-bath coupling; this is contained not in the parameter $g_{\rm q}$, but in the damping parameters $\gamma_{\rm ph}$ and $\gamma_{\rm e}$. Technically, our treatment is entirely consistent with the requirements of the Lindblad formalism and no artifacts will arise. Therefore, having a large QEPC and a small damping rate in the same system is not a contradiction. Moreover, the breakdown of the lattice is still governed by the Lindemann criterion~\cite{Lindemann1910} through the connection between allowed phonon damping and weak laser amplitude in our model. 
				
				The corresponding time evolution of phonon occupation and electron number is also shown in Figs.~\ref{fS4_SM}(b) and~\ref{fS4_SM}(c) for two phonon frequencies. The phase shift analysis due to the presence of QEPC in the previous section is also valid here for $\omega \neq \omega_0$. For $\omega = \omega_{\rm c}$ and sufficiently large QEPC~($g_{\rm q}/\omega_0 = 2$), phonon occupation first rises to diverge but it pulls down again to another NESS plateau~(largest value). At the same time, the electron number drops down to the lowest NESS plateau, as shown in the inset panels. The appearance of divergence spectrograms -- first-order phase transition -- at large QEPC is largely independent of the phonon frequency. 	\red{Finally, in Fig.~\ref{fS53_SM} we show that the phase transition occurs for other sets of parameters, as expected from Eq.~\eqref{eq_S13}, and it is a global dynamical feature of the system based on our dissipation model such that the cusp appears with varying $g_{\rm q}$ or $\omega$.}

			}
\end{document}